\documentclass[acmsmall,screen]{acmart}

\AtBeginDocument{%
  \providecommand\BibTeX{{%
    \normalfont B\kern-0.5em{\scshape i\kern-0.25em b}\kern-0.8em\TeX}}}

\setcopyright{acmcopyright}
\acmDOI{XXXXXXX.XXXXXXX}

\acmJournal{TECS}

\usepackage{tikz}
\usepackage{amsmath}
\usepackage[utf8]{inputenc}
\usepackage{textcomp}

\usepackage[normalem]{ulem}
\usepackage{setspace}
\usepackage{epsfig}
\usepackage{wrapfig}
\usepackage{multicol}
\usepackage{placeins}  
\usepackage{multirow}
\usepackage{color}
\usepackage{kotex}
\usepackage{psfrag}
\usepackage{algpseudocode}
\usepackage{multirow}
\usepackage{adjustbox}
\usepackage{courier}
\usepackage{listings}
\usepackage{pifont}
\usepackage{mathtools}
\usepackage{graphicx, pifont}
\usepackage{comment} 
\usepackage{subcaption}
\usepackage{booktabs}
\newsavebox{\largestimage}

\usepackage{filecontents}

\definecolor{orange}{RGB}{255,127,0}

\newcommand{\squishlist}{
\begin{list}{$\bullet$}
	{ \setlength{\itemsep}{0pt}      \setlength{\parsep}{-0pt}
		\setlength{\topsep}{4pt}       \setlength{\partopsep}{0pt}
		\setlength{\listparindent}{-2pt}
		\setlength{\itemindent}{-5pt}
		\setlength{\leftmargin}{1em} \setlength{\labelwidth}{0em}
		\setlength{\labelsep}{0.5em} } }

\newcommand{\squishend}{
\end{list}  }
\usepackage{framed} 
\usepackage{tcolorbox}

\newcommand{\cstore}{{\textit{\scshape CsdStore}\normalfont}}
\newcommand{\cplan}{{\textit{\scshape CsdPlan}\normalfont}}

\usepackage{kotex}
\usepackage{gensymb}
\usepackage{amsthm}
\usepackage{exscale}
\usepackage{textcomp}		
\usepackage{nccmath}
\usepackage{wrapfig}
\usepackage{mathtools}
\usepackage{xcolor}
\usepackage{color, colortbl}
\usepackage[linesnumbered,ruled,vlined]{algorithm2e}
\usepackage[normalem]{ulem}

\usepackage{lipsum}
\usepackage{setspace}
\usepackage{lscape}
\usepackage{float}
\usepackage{array}
\usepackage{tabulary}
\usepackage{multirow}
\usepackage{ctable}
\usepackage{booktabs}
\usepackage{tabularx}
\usepackage{comment}
\usepackage{aliascnt}
\newaliascnt{eqfloat}{equation}
\newfloat{eqfloat}{h}{eqflts}
\floatname{eqfloat}{Equation}

\newcommand*{\ORGeqfloat}{}
\let\ORGeqfloat\eqfloat
\def\eqfloat{%
  \let\ORIGINALcaption\caption
  \def\caption{%
    \addtocounter{equation}{-1}%
    \ORIGINALcaption
  }%
  \ORGeqfloat
}

\usepackage{mwe}
\usepackage{ragged2e}
\usepackage{tablefootnote}
\usepackage[T1]{fontenc}
\usepackage[utf8]{inputenc}
\usepackage{babel}
\usepackage[font=small,labelfont=bf]{caption}

\begin{document}

\title{An Analytical Model-based Capacity Planning Approach for Building CSD-based Storage Systems}

\author{Hongsu Byun}
\affiliation{
  \institution{Dept. of Computer Science and Engineering, Sogang University}
  \country{Seoul, South Korea}}
\email{byhs@sogang.ac.kr}

\author{Safdar Jamil}
\affiliation{
  \institution{Dept. of Computer Science and Engineering, Sogang University}
  \country{Seoul, South Korea}}
\email{safdar@sogang.ac.kr}

\author{Jungwook Han}
\affiliation{
  \institution{Dept. of Computer Science and Engineering, Sogang University}
  \country{Seoul, South Korea}}
\email{immerhjw@sogang.ac.kr}

\author{Sungyong Park}
\affiliation{
  \institution{Dept. of Computer Science and Engineering, Sogang University}
  \country{Seoul, South Korea}}
\email{parksy@sogang.ac.kr}

\author{Myungcheol Lee}
\affiliation{
  \institution{Electronics and Telecommunications Research Institute}
  \country{Daejeon, South Korea}}
\email{mclee@etri.re.kr}

\author{Changsoo Kim}
\affiliation{
  \institution{Electronics and Telecommunications Research Institute}
  \country{Daejeon, South Korea}}
\email{cskim7@etri.re.kr}

\author{Beongjun Choi}
\affiliation{
  \institution{Electronics and Telecommunications Research Institute}
  \country{Daejeon, South Korea}}
\email{bjchoi92@etri.re.kr}

\author{Youngjae Kim}
\affiliation{
  \institution{Dept. of Computer Science and Engineering, Sogang University}
  \country{Seoul, South Korea}}
\email{youkim@sogang.ac.kr}

\renewcommand{\shortauthors}{H. Byun et al.}

\begin{abstract}

The data movement in large-scale computing facilities (from compute nodes to data nodes) is categorized as one of the major contributors to high cost and energy utilization. To tackle it, in-storage processing (ISP) within storage devices, such as Solid-State Drives (SSDs), has been explored actively. 
The introduction of computational storage drives (CSDs) enabled ISP within the same form factor as regular SSDs and made it easy to replace SSDs within traditional compute nodes. With CSDs, host systems can offload various operations such as search, filter, and count. 
However, commercialized CSDs have different hardware resources and performance characteristics. 
Thus, it requires careful consideration of hardware, performance, and workload characteristics for building a CSD-based storage system within a compute node. 
Therefore, storage architects are hesitant to build a storage system based on CSDs as there are no tools to determine the benefits of CSD-based compute nodes to meet the performance requirements compared to traditional nodes based on SSDs. 
In this work, we proposed an analytical model-based storage capacity planner called \cplan{} for system architects to build performance-effective CSD-based compute nodes. 
Our model takes into account the performance characteristics of the host system, targeted workloads, and hardware and performance characteristics of CSDs to be deployed and provides optimal configuration based on the number of CSDs for a compute node.
Furthermore, \cplan{} estimates and reduces the total cost of ownership (TCO) for building a CSD-based compute node.  
To evaluate the efficacy of \cplan{}, we selected two commercially available CSDs and 4 representative big data analysis workloads. 

\end{abstract}



\maketitle
\section{Introduction}
\label{sec:intro}

\begin{figure}[t]
\centering
	\subfloat[Traditional approach]{%
    \vspace{-7pt}
    \includegraphics[width=0.32\linewidth]{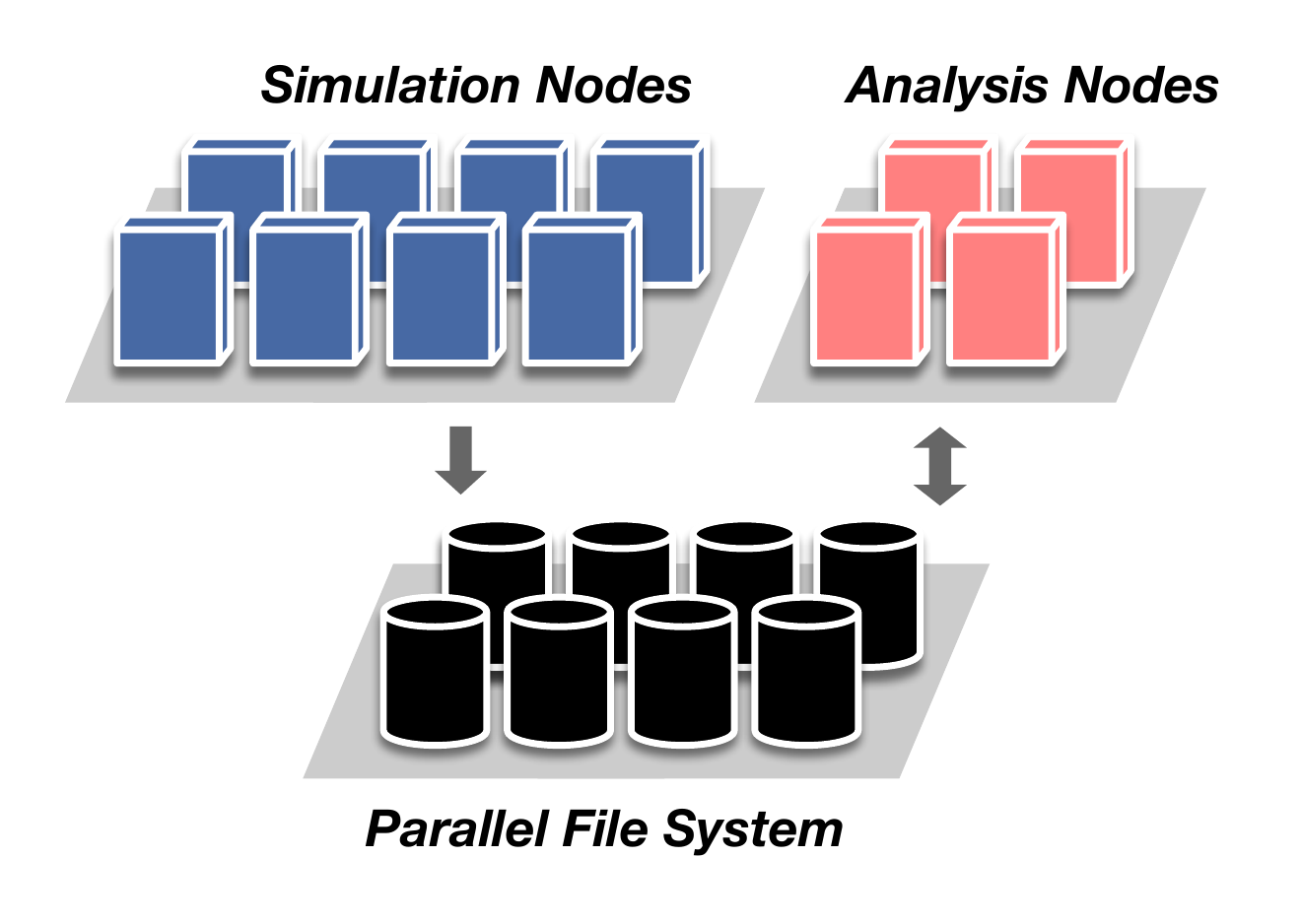}}
	\hfil
    \subfloat[In-transit approach]{%
    \vspace{-7pt}
       \includegraphics[width=0.32
       \linewidth]{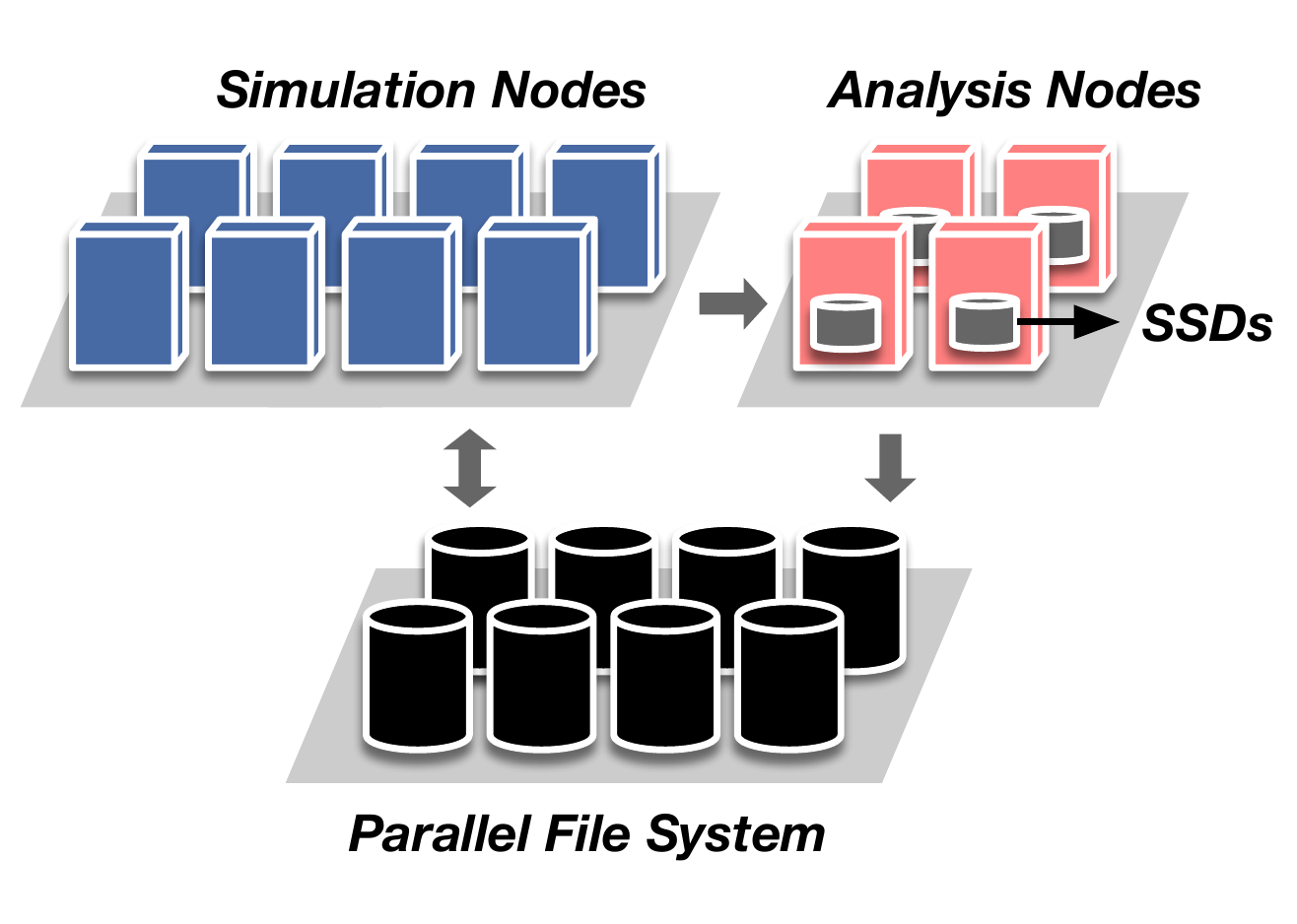}}
    \hfil
	\subfloat[In-situ approach using CSD]{%
    \vspace{-7pt}
       \includegraphics[width=0.32
       \linewidth]{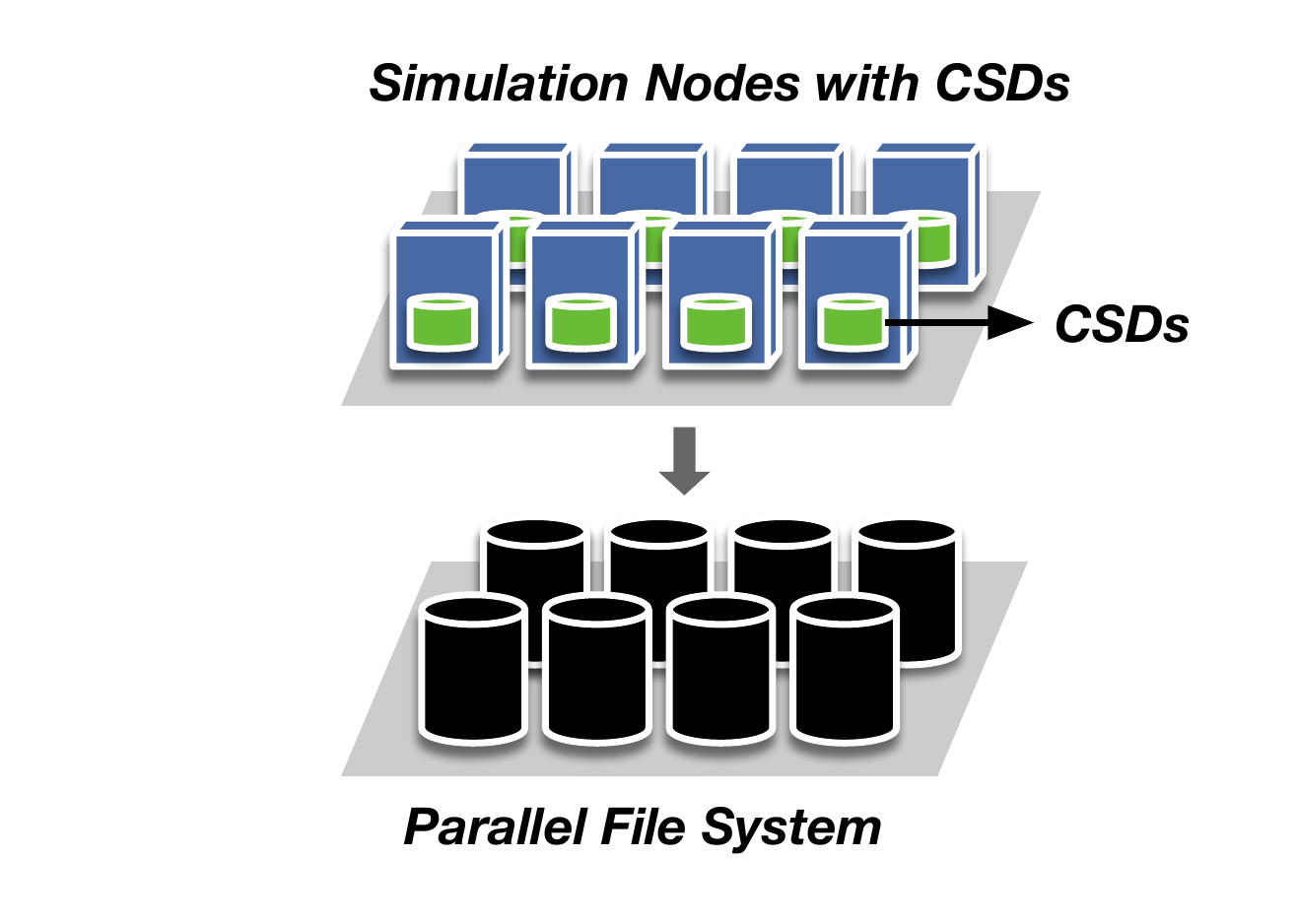}}
    \hfill
	\caption{
Three approaches for HPC workload processing: (a) traditionally using simulation node and separate analysis node, (b) in-transit approach where analysis nodes play a role of staging area and perform data analysis tasks, (c) integrating CSD into simulation node instead of analysis node~\cite{active_flash}.   }
	\label{plot:approach}
\end{figure}

High-performance computing (HPC) simulations on large-scale supercomputers (e.g., the exascale Frontier machine~\cite{frontier}, No. 1 on the Top500 list as of December 2022) routinely produce vast amounts of result output data~\cite{top500}. Examples of such applications include astrophysics, climate, combustion, and fusion. The data generated from these applications are managed by a parallel file system (PFS) such as Lustre~\cite{lustre}, as shown in Figure~\ref{plot:approach}(a). Deriving insights from output data stored at PFS often involves performing a sequence of \textit{data analysis tasks}. The data analysis tasks are performed either by a single server or a small cluster (Analysis nodes in Figure~\ref{plot:approach}(a)) in an offline manner. The critical attributes required by these tasks include parallel I/O for high performance in accessing the data from storage systems. However, these tasks suffer from huge data movement costs, leading to both performance and energy inefficiencies. 

To overcome this, a few solutions have been proposed to perform data analysis on a set of dedicated analysis nodes, where in-transit output data is analyzed before being written to the PFS, as shown in Figure~\ref{plot:approach}(b)~\cite{zheng2010predata}. Although it reduces the redundant I/Os but might cause interference at the simulation nodes\footnote{Simulation nodes are the compute nodes, and we will use these terms interchangeably from here after.} which leads to slowing down the simulation jobs. {Importantly, it still suffers from massive data transfer between the simulation node and the analysis node.}  
Therefore, HPC facilities have started looking at the potential of adopting storage devices within the simulation nodes, which provides an opportunity for adopting in-storage processing solutions~\cite{khan2021analysis}. 
In-Storage Processing (ISP) is one of the state-of-the-art paradigms that use internal resources (e.g., CPU, FPGA, and DRAM) to run data analysis tasks inside a storage device~\cite{ndp, ispdb, msst_smartssd, catalina, active_flash}. 
ISPs not only improve the energy efficiency of the system but also reduce the data movement between the host and storage devices.
A prime example of a commercially available ISP is the Computational Storage Drive (CSD). 
Recently, SK Hynix and Los Alamos National Laboratory (LANL) have demonstrated the world's first Key-Value Computational Storage Device (KV-CSD) to accelerate data analysis tasks of HPC simulations~\cite{kvcsd}. 

Moreover, several vendors have introduced commercial CSDs, including Samsung's SmartSSD~\cite{SmartSSD}, NGD system's Newport CSD~\cite{newport}, and ScaleFlux's Computational Storage~\cite{scalefulx}. The adoption of CSD within simulation nodes will play a vital role in analysis nodes where data analysis tasks can be offloaded to CSDs. Figure~\ref{plot:approach}(c) shows a representative HPC system where each simulation node has local CSD(s).
However, adopting CSDs naively does not benefit due to the distinct hardware and performance characteristics of commercially available CSDs. A typical hardware architecture of a CSD embeds an accelerator (FPGA) or an embedded CPU within a storage device to perform analysis tasks. CSDs can be classified based on the support of the operating system on top of the device. For instance, Newport SSD of the NGD systems~\cite{newport} runs an embedded OS, whereas SmartSSD~\cite{SmartSSD} does not. A CSD with built-in support of an embedded OS runs the analysis task from user space and benefits from the ease of programmability and manageability through exploiting the traditional features of OS, such as supported libraries, multitasking, and well-defined hardware abstractions.

On the other hand, a CSD, without OS support, benefits from executing the analysis kernel directly on the FPGA accelerator, just like a bare-metal application, and avoiding the software overhead caused by OS. The kernel developers with these CSDs can design and develop their analysis kernels with the supported platform. For instance, an analysis kernel for SmartSSD~\cite{SmartSSD} to be executed on the FPGA accelerator is implemented using OpenCL programming at the Vitis platform~\cite{vitis_platform} provided by Xilinx. CSDs are packaged in the same form factor as regular SSDs and can easily replace traditional block-based SSDs.
Several works~\cite{catalina, torabzadehkashi2019accelerating, torabzadehkashi2019computational} attempted to build a CSD-based storage system and executed big data applications using CSDs and showed their performance benefits. However, the performance characteristics of CSDs vary from vendor to vendor thus, adopting CSDs becomes a challenging task for storage architects within HPC facilities. For instance, storage architects have to decide whether to adopt CSDs with an embedded ARM processor (Newport CSD) or an FPGA-based accelerator (SmartSSD), as both CSDs have different computational power and programming interfaces. Moreover, the internal and external I/O bandwidth vary significantly depending on the interconnect and network protocol implementation (for more details, refer to Section~\ref{sec:motiv}). Furthermore, the performance efficiency of CSDs is highly dependent on the nature of the workload (from being compute-intensive to I/O-intensive). Previous works put strenuous efforts into identifying the optimal number of CSDs to meet performance requirements for specific workloads~\cite{catalina, torabzadehkashi2019accelerating}, thus, making the adoption of CSDs even more challenging.  

Therefore, in this work, we propose a model-based storage capacity planning tool (\cplan{}) which allows storage designers/architects to find the break-even point (BEP) effectively without having to run experiments on all storage combinations manually.  To the best of our knowledge, this work is the first to find workload-specific BEP using real commercial CSDs.
This paper makes the following specific contributions.
\squishlist
\item
We propose an analytical model-based capacity planner for \cstore{}, called \cplan{}. \cplan{} finds the optimal number of CSDs in \cstore{}, where \cstore{} outperforms a traditional approach. Our model can be extended and adopted for large-scale systems with multiple compute nodes over the network. 

\item
We developed a mathematical model for the \cplan{}, which takes the performance characteristics of the CSD and the workload patterns of the applications as input and provides the optimal number of CSDs required to outperform the traditional compute node with SSDs. The \cplan{} can give the "rule-of-thumb" to storage architects/administrators of \cstore{} while making storage capacity planning decisions.
 
\item
We performed an extensive evaluation of our proposed \cplan{} to account for various hardware characteristics, such as computing power and I/O bandwidth with several real-world workloads. Specifically, \cplan{} finds the optimal break-even point (BEP) for big data analysis workloads. For instance, the BEP for Vector Addition workload would be 5 for SmartSSDs, while it would be 1 if the computing power of the FPGA accelerator is significantly improved, likee $5\times$. 

\item 
We studied the total cost of ownership (TCO) savings in \cstore{} using \cplan{}. For example, in the Array Merge workload, \cplan{} suggests that the traditional compute node with 12 SSDs and the \cstore{} with 2$\times$ slower CPU and 6 Newport CSDs have the same throughput. According to our CSD pricing assumption, \cstore{} can reduce CPU and storage costs by up to 55\%, compared to a traditional approach. 
\squishend

This paper is organized as follows. Section~\ref{sec:back} introduces the background knowledge of CSD and the motivation for our proposed \cplan{}. Section~\ref{sec:cstore_cplan} shows an overview of \cstore{} and the mathematical analysis model of \cplan{}, and Section~\ref{sec:eval} analyzes the performance characteristics of CSD and extensive evaluation results of \cplan{}. Finally, Section~\ref{sec:conc} gives the conclusion. 
\section{Background and Motivation}
\label{sec:back}

In this section, we present the background of computational storage drives (CSDs), the related work for storage capacity provisioning, and the motivation for this study.

\subsection{Computational Storage Drives}

In-Storage Processing (ISP) uses the SSD's internal hardware resources such as CPU and memory for out-of-core execution inside the SSD~\cite{insider, Satoru, aquoman, dongup, kang2013smartssd, gu2016biscuit, koo2017summarizer, polardb, cognitivessd, jun2015bluedbm}.
The ISP not only frees up CPU and memory resources on the host, but also reduces the cost of moving data between the host and the device.
SSDs that support ISPs are called CSDs.
In the meantime, there have been many studies on the design and performance optimization of CSDs.
Biscuit~\cite{gu2016biscuit} facilitates development by defining protocols for near-data processing using the ISP and supporting a full-featured standard library and the latest C++ standards.
Willow~\cite{willow} has extended the meaning of SSD to a function that applications can use without damaging the file system by adding a programmable feature to the storage device. 
Specifically, several researches investigated to accelerate database applications with ISP on SSDs. 
FCAccel~\cite{Satoru} integrated a column-oriented field-programmable-gate-array-based acceleration engine into an SSD to offload SQL operators to the SSD.  
Aquoman~\cite{aquoman} is an SSD with a general analytic query processor, prototyped in an FPGA for ISP on the SSD. 
Smart SSD~\cite{kang2013smartssd} provides MapReduce~\cite{dean2008mapreduce} framework that can execute a user's customized job or database query inside SSD.

\renewcommand{\arraystretch}{1.3}
\begin{table}[t]
\centering
    \caption{Hardware Specifications of Representative CSDs.
    }
	\vspace{-10pt}
    \resizebox{0.8\textwidth}{!}{
\begin{tabular}{|c||c|c||c|c||c|}
\hline
\begin{tabular}[c]{@{}c@{}}\textbf{} \textbf{}\end{tabular} & \multicolumn{1}{c|}{\begin{tabular}[c]{@{}c@{}}ISP Engine\end{tabular}} & 
\multicolumn{1}{c|}{\begin{tabular}[c]{@{}c@{}}External I/O BW\end{tabular}} & \multicolumn{1}{c|}{\begin{tabular}[c]{@{}c@{}}Internal I/O BW\end{tabular}}  \\
\hline\hline
Smart SSD\tablefootnote{Here, Smart SSD is the research prototype name of the cited paper, which is different from SmartSSD, a commercial CSD.}~\cite{smartssds} &
\multicolumn{1}{c|}{ARM Cortex R4 @ 400~MHz} & 
\multicolumn{1}{c|}{0.55~GB/s}  & \multicolumn{1}{c|}{1.5~GB/s} \\ \hline
Willow~\cite{willow} & \multicolumn{1}{c|}{FPGA @ 125~MHz} & 
\multicolumn{1}{c|}{2~GB/s}  & \multicolumn{1}{c|}{4~GB/s} \\ \hline
Biscuit~\cite{biscuit} & \multicolumn{1}{c|}{ARM Cortex R7 @ 750~MHz} & 
\multicolumn{1}{c|}{3.2~GB/s}  & \multicolumn{1}{c|}{4~GB/s} \\ \hline
\end{tabular}
}\label{tbl:csd_spec}
\end{table}

\subsection{Storage Capacity Provisioning}

There have been several storage capacity provisioning studies that cost-effectively design compute nodes using SSDs instead of HDDs~\cite{eurosys09, hybridstore, hybridplan, dbaas}. 
Among these, especially Narayanan et al.~\cite{eurosys09} investigated the role of SSDs in enterprise compute nodes using multiple real-world data-center traces. 
Their work explores the cost-benefit trade-offs of various SSD and HDD configurations. 
Y. Kim et al.~\cite{hybridstore} investigated the problem of finding the optimal storage configuration for compute nodes employing both SSDs and HDDs while meeting performance requirements. They also studied the issue of designing the dynamic placement of workload in the hybrid storage configurations. 
On the other hand, our work is different from these works. 
We explored the problem of cost-benefit trade-offs of various CSD configurations for big data workloads.
Since CSDs are much more complex devices than SSDs, 
finding the design requirements, such as the number of CSDs when building the CSD-array-based computational node is not an easy task.
We explored the computational and I/O processing performance of commercial CSDs. We found that (i) the performance spectrum of CSDs is very broad (not comparable to SSDs), and (ii) even CSDs have different performance trends. More details on these are given in the next section.

\subsection{Motivation}
\label{sec:motiv}

Table~\ref{tbl:csd_spec} shows the hardware specifications of representative CSDs.
In Table~\ref{tbl:csd_spec}, all CSDs have different computing engines (e.g., FPGAs or low-power CPUs), but they all have much higher internal I/O bandwidth than external I/O bandwidth. 
However, the real performance of the kernel of applications is dependent on the workload patterns, that is, depending on whether the kernel is compute-intensive or I/O-intensive.
Additionally, the application's I/O bandwidth also depends on whether it is external I/O or internal I/O.
To verify these, we conducted several experiments and measured the execution times and I/O bandwidths over two commercial CSDs (SmartSSD and Newport CSD). 

\begin{figure}[!t]
\centering
	\subfloat[Execution Time]{%
       \includegraphics[width=0.4\linewidth]{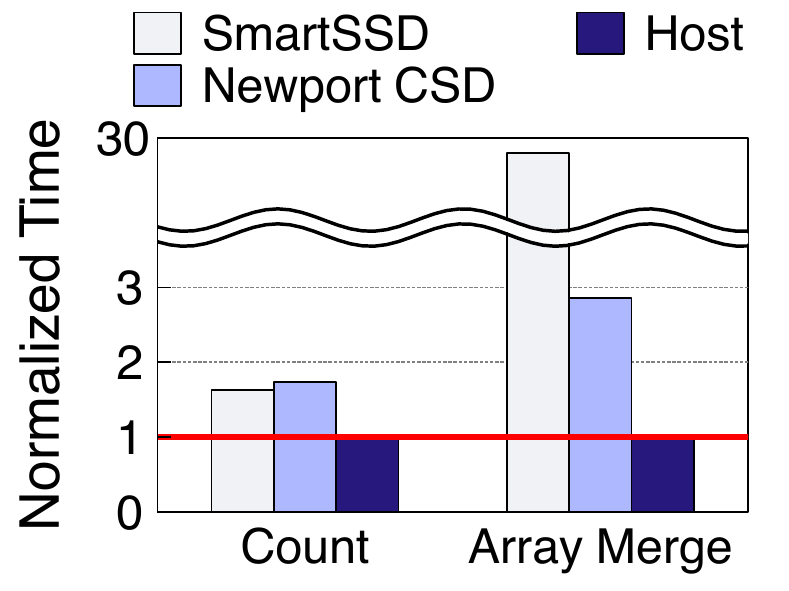}}
	\hfil
	\subfloat[I/O Bandwidth]{%
       \includegraphics[width=0.4\linewidth]{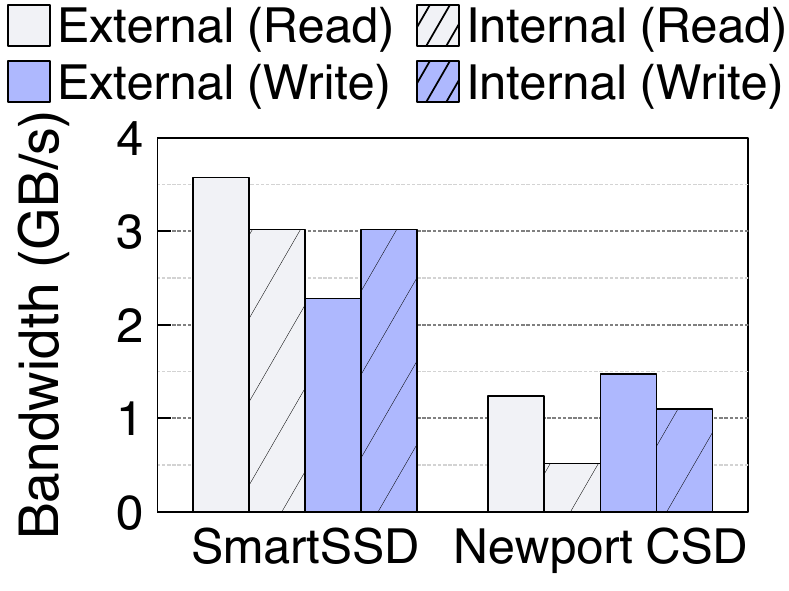}}
    \hfill
	\caption{
(a) Comparison of execution times of Count and Array Merge workloads against CSDs and host CPU. Execution time is normalized to the execution time of the host system.
(b) Comparison of internal and external I/O bandwidths of CSDs.
    }
	\label{plot:csd_comp_bw}
\end{figure}

Figure~\ref{plot:csd_comp_bw} shows the results of the execution times for two represented analysis kernels, Array Merge and Count. 
A detailed description of the analysis kernels and the input data size are provided in Section~\ref{sec:eval}. 
{In Figure~\ref{plot:csd_comp_bw}(a), the execution times are normalized with the host system using a single CPU core and equipped with traditional SSD.}
For Count, we observe that SmartSSD exhibits a slightly lower execution time than Newport CSD. 
Also, we observe that the execution times of SmartSSD and Newport CSD were about 1.6$\times$ and 1.7$\times$ worse than that of the host system, respectively.
Through this, it can be seen that the performance difference between FPGA and ARM processors is small for the Count kernel. 
On the other hand, {for Array Merge, }we observe that the execution times of SmartSSD and Newport CSD were 28$\times$ and 2.8$\times$ higher than that of the host, respectively. 
SmartSSD showed significantly higher execution time compared to the host as well as the Newport CSD.
In the Array Merge kernel, the reason why CSD has a much higher execution time than the host system is that Array Merge is a more CPU-bounded workload than Count and requires a system with high computational power.
On the other hand, CSD is a system with significantly lower computational power than the host CPU. 
Since the Count kernel is an I/O-intensive workload, the execution time is not significantly different from that of the host system.

Next, we measured the internal and external bandwidths of Newport CSD and SmartSSD. 
To measure the external I/O bandwidth of each CSD, we ran the FIO benchmark~\cite{fio} on the host.
However, due to the difference in hardware design between SmartSSD and Newport CSD, the internal I/O bandwidth measurement method is as follows: For the Newport CSD, to measure internal and external I/O bandwidth, we performed an FIO benchmark~\cite{fio} on the host and CSD side. The FIO benchmark was configured by using the libaio engine\footnote{Linux-native asynchronous I/O access library}, direct option on, 1~MB request size, 64 queue depth, and sequential pattern.
For the SmartSSD, for internal I/O bandwidth measurement, we used a bandwidth measurement kernel program~\cite{p2p_bandwidth} with a request size of 64~MB.
Figure~\ref{plot:csd_comp_bw}(b) shows the I/O bandwidth measurements for SmartSSD and Newport CSD.
We define $R_{\mathrm{tx}}$ as the ratio of the {internal I/O bandwidth} to the {external I/O bandwidth} ($R_{\mathrm{tx}}{=}\frac{BW_{\mathrm{Internal}}}{BW_{\mathrm{External}}}$).
If $R_{\mathrm{tx}}$ is greater than 1, it means that the internal I/O bandwidth is higher than the external I/O bandwidth. 

In the results for SmartSSD, we observe that for reads, the external bandwidth is about 1.18$\times$ higher than the internal bandwidth ($R_{\mathrm{tx}}{=}0.84$), but, for writes, the internal bandwidth is about 1.36$\times$ higher than the external bandwidth ($R_{\mathrm{tx}}{=}1.36$). This is due to the hardware limitations of the SmartSSD, the maximum internal bandwidth is bound to the bandwidth of the PCIe bus connecting the SSD and the FPGA.
In the results for Newport CSD, 
we observe, surprisingly, that the external bandwidth is 2.28$\times$ and 1.33$\times$ higher than the internal bandwidth for both read and write workloads, respectively ($R_\mathrm{tx}{=}0.43,\; 0.75$). 
{The NGD system explained that the internal components of Newport CSD (DRAM, NAND, Etc.) are connected through high-speed interconnect but did not disclose detailed hardware specifications~\cite{newport-paper}.
}
These results show a different observation from the observations in the literature where the internal I/O bandwidth of CSD is higher than the external I/O  bandwidth~\cite{biscuit, active_flash, smartssds, willow}.

To summarize our observations, each CSD shows different performances depending on their computing power, I/O bandwidth, and workload characteristics. Therefore, when building a compute node based on a CSD-array, storage architects should design in careful consideration of each SSD's hardware and workload characteristics. 
Therefore, this study aims to offer a software tool that provides storage architects with guidance when building the CSD-array-based node in terms of the optimal number of CSDs for CSD-based compute nodes. 

\section{Capacity Planning for \cstore{}}
\label{sec:cstore_cplan}
This section presents an overview of \cstore{} and details of how to build a \cplan{} for \cstore{} and how system architects/administrators can use it.

\subsection{Overview of \cstore{}}

\begin{figure}[t]
	\centering
    \includegraphics[width=0.8\linewidth]{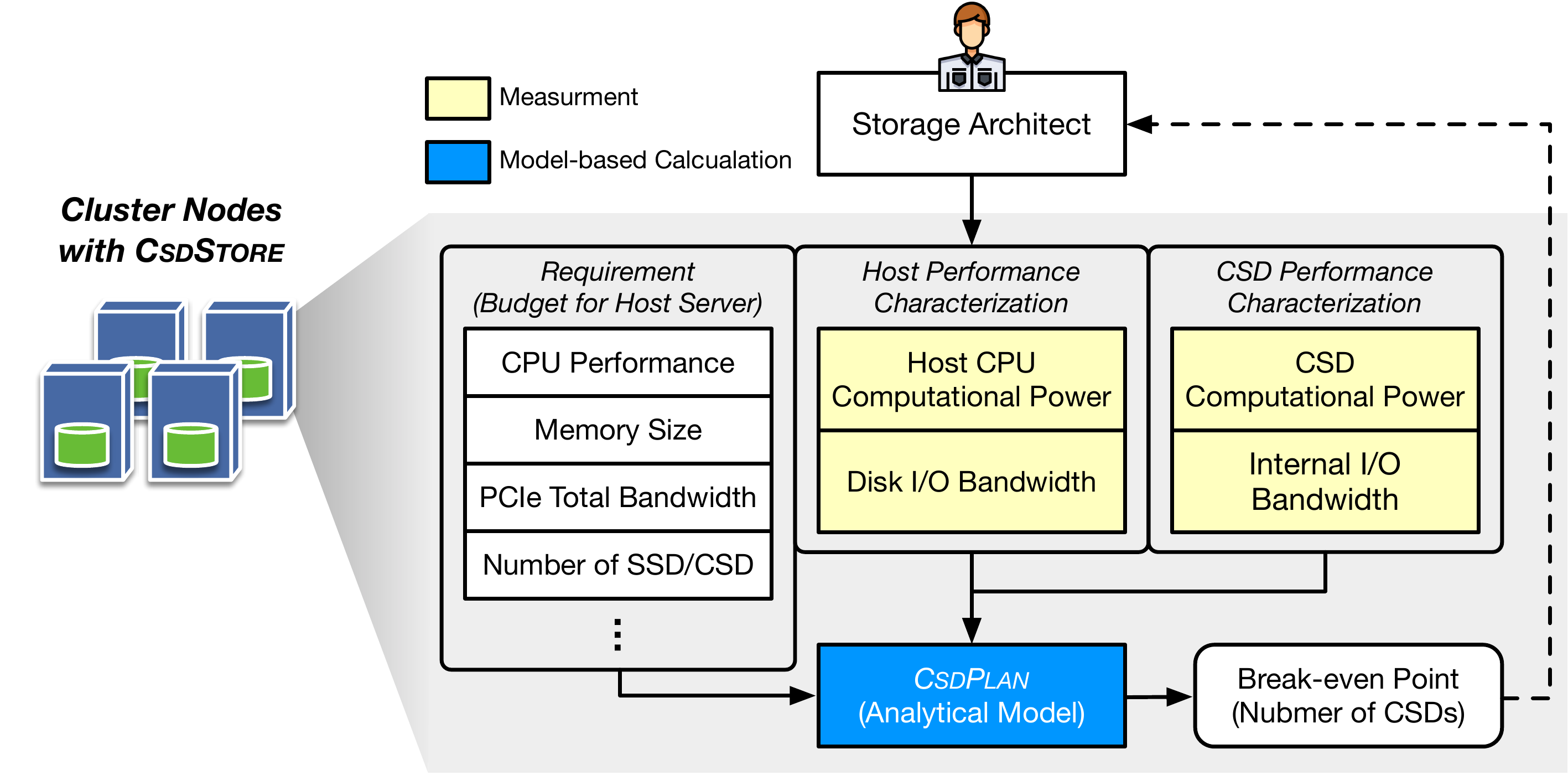}
	\caption{
    {{\cstore{}} and \cplan{}'s overview description. Yellow boxes indicate values that system architects should measure.} 
	}
    \vspace{-5pt}
	\label{pics:csdplan}
\end{figure}

\cstore{} is a cluster of CSDs and leverages the compute capabilities of each CSD to provide better performance than the traditional storage server approach. 
Several recent studies have shown the potential of CSD-array for HPC, big data and AI workloads~\cite{torabzadehkashi2019accelerating, torabzadehkashi2019computational, newport-paper}.
However, one of the main problems for system architects with CSD-arrays is the lack of planning tools for how to efficiently build CSD-array based on their requirements. For instance, with varying computation resources, system architects are not able to identify the number of CSDs to be installed in a CSD-array. Thus, it is possible that a system architect might over- or under-provision the CSD-array's resource when designing a cluster.

\cstore{} provides \cplan{} as a software tool to the system architects. 
{\cplan{}, which provides guidelines on how to efficiently build a compute node using CSDs, on a limited budget.}
We envision \cplan{} to be a tool that would enable system architects to provision the CSD-array based computational node in performance-effective ways. 
Figure~\ref{pics:csdplan} depicts an overview of the \cplan{} where a system architect is required to measure the initial performance characteristics of the host and CSDs. 
\cplan{} employs a break-even point (BEP) decision-maker based on mathematical formulations to make its storage provisioning decisions. 
\cplan{} finds the BEP where the performance of data analysis in a CSD-array configuration outperforms the traditional storage server approach. 
{system architects can understand the effectiveness of the CSD-array system by considering both the BEP found with \cplan{} and the budget to be used for the storage server implementation.} {For example, with \cplan{}, a \cstore{} with higher performance than a traditional storage server can be built at a lower cost.}

\subsection{\cplan{}: Capacity Planning}
\label{sec:csdllan}
\cplan{} is a software module that provides guidelines to system architects when building a computational node.
\cplan{} takes input as the computation and I/O performance parameters of a CSD for applications and outputs the minimum number of required CSDs for \cstore{}. 
\cplan{} required the following two steps to be performed by the system architect:
\squishlist
\item
\textbf{Step 1}: The system architect selects the CSD to be deployed at the CSD-array and targets applications that will be running on that cluster. As the \cplan{} takes the performance characteristics as input, thus a system architect is required to measure the performance of the CSD to obtain the computational and I/O processing capabilities of the corresponding CSD. 

\item
\textbf{Step 2}: Once the performance characteristics are obtained from Step 1, the system architect inputs them to the \cplan{}, which determines the minimum number of CSDs required to achieve optimal performance on a CSD-array that is higher than the traditional storage server approach.
\squishend
Our \cplan{}'s solver is built on top of mathematical system modeling and is described in detail in the following subsection.

\subsubsection{System Modeling} 
In this subsection, we provide details of performance modeling of \cplan{} for SSD system and CSD system as follows: 
We first define a system equipped with a single device: 

\squishlist
\item {\textbf{SSD system ($n$)}}: A traditional compute node where a host is equipped with a single block-based SSD and uses the host $n$ cores and memory for analysis. 

\item {\textbf{CSD system}}: {A compute node equipped with commercial CSDs where utilizes the CSD's resources for analysis. 
} 

\squishend
The execution time is considered the performance metric in our model.

The execution time of the SSD system ($n$) ($T_{\mathrm{SSD}(n)}$) for the kernel ($W$) can be modeled as the sum of the data transfer time ($T_{\mathrm{SSD}\text{-}\mathrm{tx}}$) and the computation time ($T_{\mathrm{SSD}(n)\text{-}\mathrm{comp.}}$). 
Thus, 
\begin{align}
T_{\mathrm{SSD}(n)}=T_{\mathrm{SSD}\text{-}\mathrm{tx}} + T_{\mathrm{SSD}(n)\text{-}\mathrm{comp.}}
\end{align}

We assume that the host system is comprised of $n$ cores and the workload is equally divided between all the cores. 
The data transfer time stays the same regardless of the number of cores in the host system. 
The sequential execution time is presented as $T_{\mathrm{SSD}\text{-}\mathrm{comp.}}$, thus according to Amdahl's law~\cite{Gustafson2011}, the execution time over $n$ cores would be: 
\begin{align}
T_{\mathrm{SSD}(n)\text{-}\mathrm{comp.}}=\frac{1}{n} {{\cdot}} T_{\mathrm{SSD}\text{-}\mathrm{comp.}} ,\; n \le \text{Max Cores}
\end{align}

On the other hand, the performance model for the CSD system will be:
\begin{align}
T_\mathrm{CSD}=T_\mathrm{CSD\text{-}tx} + T_\mathrm{CSD\text{-}comp}.
\label{eq:csd_default}
\end{align}

Moreover, the resources of the host system are classified as \textit{normal} and \textit{overloaded} based on the workload. For instance, if the kernel is being executed in parallel to other applications at the host system/machine and the execution time of the kernel is greatly affected. This is due to resources being shared between the kernel and other applications and thus leading to resource contention. We call this situation an \textit{overloaded} condition. On the other hand, if the kernel is being executed with the desired resources from the host machine, it will be considered a \textit{normal} condition.

Applications running parallel to the kernel are categorized as: CPU-, data-, or memory-intensive.
If an application is CPU-intensive, than the computational resources are exhausted, thus affect the computation time of the kernel. On the other hand, if an application is data-intensive, than the I/O resources are being shared thus leading to significant increase in data transfer time. 
However, if an application is memory-intensive, then both computation and data transfer times are affected due to high I/O overhead by frequent disk swapping in the virtual memory system. 
Therefore, we extend the execution time model of each system in overloaded conditions by applying the \textit{slow-down factor} (the rate of increase in time) to the computation time and data transfer time. Assume that the slow-down factors of data transfer time and computation time are $sd_\mathrm{tx}$ and $sd_\mathrm{comp.}$, respectively.  
The execution time of the SSD system can be modeled as follows: 

\begin{align}
T_{\mathrm{SSD}(n)} = sd_\mathrm{tx} {{\cdot}} T_{\mathrm{SSD}\text{-}\mathrm{tx}} + sd_\mathrm{comp.} {{\cdot}}  T_{\mathrm{SSD}(n)\text{-}\mathrm{comp.}} 
\;\;(sd_\mathrm{tx},\; sd_\mathrm{comp.} \ge 1) 
\label{eq:ssd_overload}
\end{align}

The CSD systems, unlike SSD systems, run the kernel on CSD, so their execution time is not affected by overload conditions.
Therefore, the slow-down factor for each term of the CSD system is always 1.
Thus, the execution time of the CSD system is simply modeled as to Equation~(\ref{eq:csd_default}).

Now, we extend this performance model to a system comprised of an array of devices.
We assume that the data required for workload execution is uniformly distributed and stored in the device array.
We consider two systems as follows: 

\squishlist
\item {\textbf{SSD-array system (${n}$)} ($T_{\mathrm{SSD}(n)\text{-}\mathrm{array}}$): The host is equipped with an array of block-based SSDs and uses the host CPU's $n$ cores and memory to run analysis kernels.}

\item \textbf{CSD-array system} ($T_\mathrm{CSD\text{-}array}$): The host is equipped with an array of CSDs and uses the CSD’s CPU and memory instead of the host’s resources to run the analysis kernel.
\squishend
An array of $m$ SSDs can theoretically reduce the data transfer time to $\frac{1}{m}$ until the bus connected to the host becomes a bottleneck~\cite{spdk}.
Therefore, we set the number of SSDs ($M_\mathrm{SSD}$) as $k_\mathrm{limit}$ so that the disk I/O becomes a bottleneck in the SSD-array system. 
Thus, the SSD-array system model is extended as follows:

\begin{equation}
\begin{aligned}
    T_{\mathrm{SSD}(n)\text{-}\mathrm{array}} = \frac{1}{M_\mathrm{SSD}} {{\cdot}}  sd_\mathrm{tx} {{\cdot}} T_{\mathrm{SSD}\text{-}\mathrm{tx}} + sd_\mathrm{comp.} {{\cdot}}
    T_{\mathrm{SSD}(n)\text{-}\mathrm{comp.}} 
    \label{eq:host_array}
\end{aligned}
\end{equation}

\begin{gather}
M_\mathrm{SSD}= 
    \left\{ 
        \begin{matrix}
            m  & \mathrm{if}~ (m < k_\mathrm{limit})  \\ 
            k_\mathrm{limit}  & \mathrm{else} 
        \end{matrix}
    \right.
    \label{eq:host_array_limit}
\end{gather}

Unlike the SSD-array system, since each CSD has the computational capability, both the data transfer time and computation time of the CSD-array system are reduced to $\frac{1}{m}$. In addition, since each CSD does not share the connected bus, there is no bottleneck as the number of CSDs increases.
The CSD-array system model is extended as follows:

\begin{flalign}
    &T_\mathrm{CSD\text{-}array} = \frac{1}{M_\mathrm{CSD}} {{\cdot}} T_\mathrm{CSD\text{-}tx} + \frac{1}{M_\mathrm{CSD}} {{\cdot}}  T_\mathrm{CSD\text{-}comp.}
    \label{eq:csd_array}
\end{flalign}

\subsubsection{Solver: Finding the Break-Even Point}

\cplan{} deploys a solver to find the BEP for the number of CSDs in a CSD-array-based compute node. Our proposed solver takes the performance characteristics of the CSDs as input and generates an optimal number of CSDs as output. This optimal number of CSDs is referred to as the BEP, where the CSD-array will outperform the traditional compute node.
Therefore, we derive the mathematical model of ($T_{\mathrm{SSD}(n)\text{-}\mathrm{array}} >  T_\mathrm{CSD\text{-}array}$)  
as follows:

\begin{gather*}
T_{\mathrm{SSD}(n)\text{-}\mathrm{array}} >  T_\mathrm{CSD\text{-}array}
\end{gather*}

\begin{flalign*}
\Rightarrow \;\; &\frac{sd_\mathrm{tx}}{M} {\cdot} T_{\mathrm{SSD}\text{-}\mathrm{tx}} + sd_\mathrm{comp.} {\cdot} T_{\mathrm{SSD}(n)\text{-}\mathrm{comp.}}> 
 \frac{1}{M} {\cdot} T_\mathrm{CSD\text{-}tx} + \frac{1}{M} {\cdot} T_\mathrm{CSD\text{-}comp.} 
\end{flalign*}

\begin{flalign*}
sd_\mathrm{tx} {{\cdot}} T_{\mathrm{SSD}\text{-}\mathrm{tx}} + M {\cdot} sd_\mathrm{comp.} {\cdot} T_{\mathrm{SSD}(n)\text{-}\mathrm{comp.}} >
T_\mathrm{CSD\text{-}tx} + T_\mathrm{CSD\text{-}comp.}
\;\; \text{(Multiply both sides by $M$)} \qquad\quad\;\;\;
\end{flalign*}

\begin{flalign*}
M {\cdot} sd_\mathrm{comp.} {{\cdot}} T_{\mathrm{SSD}(n)\text{-}\mathrm{comp.}}>
T_\mathrm{CSD\text{-}tx} +  T_\mathrm{CSD\text{-}comp.} {-} sd_\mathrm{tx} {\cdot} T_{\mathrm{SSD}\text{-}\mathrm{tx}}
\;\; \text{(Subtract $sd_\mathrm{tx} {{\cdot}} T_{\mathrm{SSD}\text{-}\mathrm{tx}}$ from both sides)}
\end{flalign*}

\begin{flalign*}
M > \frac{   T_\mathrm{CSD\text{-}tx} +  T_\mathrm{CSD\text{-}comp.} - sd_\mathrm{tx} {\cdot} T_{\mathrm{SSD}\text{-}\mathrm{tx}} }{sd_\mathrm{comp.} {\cdot} T_{\mathrm{SSD}(n)\text{-}\mathrm{comp.}}} 
\;\; \text{(Divide both sides by $sd_\mathrm{comp.} {\cdot} T_{\mathrm{SSD}(n)\text{-}\mathrm{comp.}}$)} \qquad\qquad\quad\;\;
\end{flalign*}

\begin{flalign*}
M_\textrm{CSD} > \frac{T_\mathrm{CSD\text{-}tx} +  T_\mathrm{CSD\text{-}comp.} - sd_\mathrm{tx} {\cdot} T_{\mathrm{SSD}\text{-}\mathrm{tx}} }{sd_\mathrm{comp.} {\cdot} T_{\mathrm{SSD}(n)\text{-}\mathrm{comp.}}} 
\;\;
\text{(To find BEP, $M=M_\textrm{SSD}=M_\textrm{CSD}$)} \qquad\qquad\qquad\qquad\;
\end{flalign*}

Therefore,

\begin{flalign}
 M_\mathrm{CSD} \ge \left\lceil \frac{T_\mathrm{CSD\text{-}tx} +  T_\mathrm{CSD\text{-}comp.} - sd_\mathrm{tx} {\cdot} T_{\mathrm{SSD}\text{-}\mathrm{tx}} }{sd_\mathrm{comp.} {\cdot} T_{\mathrm{SSD}(n)\text{-}\mathrm{comp.}}} \right\rceil \footnotemark
 \label{eq:model_prim}
\end{flalign}
\footnotetext{($\lceil \; \rceil$) is least integer function}

If the host resource is not {overloaded}, $sd_{tx}=1$, $sd_{comp.}=1$, then the following holds.

\begin{flalign}
M_\mathrm{CSD} \ge \left\lceil \frac{T_\mathrm{CSD\text{-}tx} +  T_\mathrm{CSD\text{-}comp.} - T_{\mathrm{SSD}\text{-}\mathrm{tx}} }{T_{\mathrm{SSD}(n)\text{-}\mathrm{comp.}}} \right\rceil 
\end{flalign}

{\bf Impact of Computing and I/O Performance}: The increase or decrease of the computational power of CSD determines the change in kernel execution time. Additionally, the internal I/O bandwidth of the CSD can be higher or lower than the external I/O bandwidth. CSD's computing power and internal I/O bandwidth are determined by how the device is manufactured. 
In Table~\ref{tbl:csd_spec}, the internal I/O bandwidth of CSD is higher than the external I/O bandwidth. 
On the other hand, as shown in Figure~\ref{plot:csd_comp_bw}(b), the internal I/O bandwidth of CSD can be lower than the external I/O bandwidth. 
Therefore, we model the BEP ($ N_\mathrm{CSD}$) according to the change of CSD's computational power and internal I/O bandwidth as follows.

\begin{equation}
\begin{aligned}
S(T_\mathrm{CSD\text{-}tx}&, \; T_\mathrm{CSD\text{-}comp.}) = \left\lceil \frac{T_\mathrm{CSD\text{-}tx} +  T_\mathrm{CSD\text{-}comp.} - T_{\mathrm{SSD}\text{-}\mathrm{tx}} }{T_{\mathrm{SSD}(n)\text{-}\mathrm{comp.}}} \right\rceil 
\label{eq:model_prim_func}
\end{aligned}
\end{equation}

To simplify the formula, we define the following ratios:

\begin{flalign*}
\;R_{(n)\mathrm{comp.}} = \frac{ T_{\mathrm{SSD}(n)\text{-}\mathrm{comp.}}}{T_\mathrm{{CSD\text{-}comp.}}}, \;R_{(n)\mathrm{SSD}} = \frac{ T_{\mathrm{SSD}\text{-}\mathrm{tx}}}{T_{\mathrm{SSD}(n)\text{-}\mathrm{comp.}}}
\end{flalign*}

Then,

\begin{flalign*}
 \frac{T_\mathrm{CSD\text{-}tx} +  T_\mathrm{CSD\text{-}comp.} - T_{\mathrm{SSD}\text{-}\mathrm{tx}} }{T_{\mathrm{SSD}(n)\text{-}\mathrm{comp.}}}  = \frac{T_\mathrm{{CSD\text{-}tx}}}{T_{\mathrm{SSD}(n)\text{-}\mathrm{comp.}}} + R^{-1}_{(n)\mathrm{comp.}} {-} R_{(n)\mathrm{SSD}}
\end{flalign*}

We assumed that the SSD system uses the CSD as a block device. 

\begin{align*}
R_\mathrm{tx} = \frac{BW_\mathrm{Internal}}{BW_\mathrm{External}} = \frac{\mathrm{Data\;Size}/BW_\mathrm{External}}{\mathrm{Data\; Size}/BW_\mathrm{Internal}} = \frac{T_\mathrm{{SSD\text{-}tx}}}{T_\mathrm{{CSD\text{-}tx}}} ,\;\; T_\mathrm{{CSD\text{-}tx}} = R^{-1}_\mathrm{tx} {\cdot} T_\mathrm{{SSD\text{-}tx}}
\end{align*}

Then,

\begin{align*}
 &\frac{T_\mathrm{{CSD\text{-}tx}}}{T_{\mathrm{SSD}(n)\text{-}\mathrm{comp.}}} + R^{-1}_{(n)\mathrm{comp.}} {-} R_{(n)\mathrm{SSD}} \\&= \frac{R^{-1}_\mathrm{tx} {\cdot} T_\mathrm{{SSD\text{-}tx}}}{T_{\mathrm{SSD}(n)\text{-}\mathrm{comp.}}} + R^{-1}_{(n)\mathrm{comp.}} {-} R_{(n)\mathrm{SSD}} \;\; \left( T_\mathrm{{CSD\text{-}tx}} = R^{-1}_\mathrm{tx} {\cdot} T_\mathrm{{SSD\text{-}tx}} \right)
 \\&=
  R^{-1}_\mathrm{tx} {\cdot} R_{(n)\mathrm{SSD}} + R^{-1}_{(n)\mathrm{comp.}} {-} R_{(n)\mathrm{SSD}} \;\; ( R_{(n)\mathrm{SSD}} = \frac{ T_{\mathrm{SSD}\text{-}\mathrm{tx}}}{T_{\mathrm{SSD}(n)\text{-}\mathrm{comp.}}})
  \\&=
  (R^{-1}_\mathrm{tx}-1) {\cdot} R_{(n)\mathrm{SSD}} + R^{-1}_{(n)\mathrm{comp.}} \;\; (\text{Distributive Law})
\end{align*}

Therefore, Equations~(\ref{eq:model_prim}) is transformed as follows. 

\begin{equation}
\begin{aligned}
\label{eq:model_func_normal-2}
     S(R_\mathrm{tx},\;R_{(n)\mathrm{comp.}}) = \left\lceil (R^{-1}_\mathrm{tx}-1) {\cdot} R_{(n)\mathrm{SSD}} + R^{-1}_{(n)\mathrm{comp.}} \right\rceil
\end{aligned}
\end{equation}

Finally, the BEP cannot be smaller than 1, so set the minimum value to 1 as follows.

\begin{equation}
\begin{aligned}
\label{eq:model_func_normal}
     S(R_\mathrm{tx},\;R_{(n)\mathrm{comp.}}) = \text{max} \left( \left\lceil (R^{-1}_\mathrm{tx}{-}1) {\cdot} R_{(n)\mathrm{SSD}} + R^{-1}_{(n)\mathrm{comp.}} \right\rceil ,1\right)
\end{aligned}
\end{equation}

By adjusting the two variables in the above function, we can estimate the change in BEP when building a CSD-array-based compute node.

\section{Evaluating \cplan{}}
\label{sec:eval}
{This section presents an evaluation of CSD and \cplan{}. To this end, in Section~\ref{sec:csd_perf}, CSD performance characteristics are first evaluated to obtain the parameters required for \cplan{} use, and then \cplan{} is evaluated from Section~\ref{sec:eval_solver}.}
\renewcommand{\arraystretch}{1.3}
\begin{table}[!b] 
    \centering
    \caption{Host server specifications.}
	\vspace{-10pt}
    \resizebox{0.85\textwidth}{!}{
       \begin{tabular}{|c||c|}
            \hline
            CPU & AMD EPYC\textsuperscript{TM} 7352,   
              24 Cores (48 Threads), 2.3~GHz (Up to 3.2~GHz) \\ \hline
            Socket & 2 NUMA Node \\ \hline
            Memory  & 256~GB (64~GB $\times$ 4) DRAM
                     DDR4 3200~MHz\\ \hline
            OS  & Centos 7.92.2009 (Core) / Linux Kernel 4.14\\ \hline
        \end{tabular}
    }
    \label{tbl:host}
\end{table}

\renewcommand{\arraystretch}{1.3}
\begin{table}[!b]
\centering
    \caption{CSD Specifications.
    }
	\vspace{-10pt}
    \resizebox{0.85\textwidth}{!}{
\begin{tabular}{|c||c|c|}
\hline
                   & \textbf{SmartSSD}~\cite{SmartSSD} &\textbf{Newport CSD}~\cite{newport} \\ \hline\hline
Storage Capacity                  & 3.84~TB        & 8~TB           \\ \hline
Host Interface                  & PCIe Gen3$\scriptstyle\times$4 (U.2)        & PCIe Gen3$\scriptstyle\times$4 (U.2)           \\ \hline
\multirow{3}{*}{\begin{tabular}[c]{@{}c@{}}In-Stroage \\ Processing \\ Engine\end{tabular}} & Xilinx Kintex Ultrascale+ KU15P        & ARM Cortex-A53 1.0~GHz, 4~Cores           \\ \cline{2-3} 
                   & 4~GB DDR4 DRAM, 4.325~MB BRAM        & 8~GB DDR4 DRAM           \\ \cline{2-3} 
                   & Clock : 300~MHz       & OS : Linux Kernel 4.14            \\ \hline
\end{tabular}
    }
\label{tbl:smartssd+newport} 
\end{table}

\subsection{Experiment Setup}
We implemented our proposed BEP solver, \cplan{}, using Python, and the source code is less than 100 lines.
\cplan{} takes an extremely short time (in terms of seconds) to find an optimal BEP for CSD-array. However, the initial evaluation of CSD for the workload may require some effort, depending on the accuracy of the CSD's performance characterization.

For evaluation, we build two systems with two AMD EPYC\textsuperscript{TM} 7352 CPUs with 24 cores and 256~GB DRAM, running CentOS 7
where each of the systems has SmartSSD and Newport CSD, respectively.
SmartSSD does not have an OS installed and runs the kernel using an FPGA accelerator.
On the other hand, Newport CSD runs a Linux-based embedded OS using an ASIC-based 64-bit general-purpose CPU and runs the kernel on it.
Detailed specifications of the host server and each CSD are shown in Table~\ref{tbl:host} and~\ref{tbl:smartssd+newport}.

To evaluate the efficiency of our proposed \cplan{} for \cstore{}, we selected four widely adopted analysis kernels from big data applications. 
The analysis kernel and their corresponding workload {working set} size (in parenthesis) are listed below:

\squishlist
\item {Count (4.8~GB):} Counts specific values in one integer array
\item {Vector Addition (4.8~GB):} Calculates the sum of each element of two integer arrays and creates one large integer array
\item {Array Merge (4.8~GB):} Takes two sorted integer arrays as input, removes duplicate elements, merges them, and creates a new merged array
\item {Page Rank (0.2~GB\footnote{{The execution time of Page Rank is extremely long and increases exponentially as the working set increases, so the working set is smaller than other workloads.}}):} Performs Page Rank algorithm~\cite{pagerank} for graph data processing using the rank values of pages stored in one float-type two-dimensional array and one float-type one-dimensional vector
\squishend

\subsection{Performance Characterization of CSDs}
\label{sec:csd_perf}

Our proposed \cplan{} relies on system architects to evaluate the performance characterization of the CSDs. Thus, in this subsection, we present the evaluation steps for the performance characterization of CSD and the process of analyzing the results for selected big data workloads.
{To this end, the throughput was shown by measuring the data transfer time and computation time when offloading the analysis kernel in CSD.
The kernel offloading overhead from the host side to CSD is considered to be not that significant in our experiment setup, thus, we did not take into account that overhead.
Data transfer time is the time for loading the working set from NAND in the CSD to memory, and computation time is the time for computing the working set loaded in the memory of the CSD.
When processing a working set of 4.8~GB, Newport CSD has 8~GB of memory, so it processes the workload with 1 I/O, and SmartSSD has 4~GB of memory, so it processes with 2 I/O.}

\subsubsection{Impact of Parallel Computation}
\label{sec:eval_smartssd}
The CSDs in our system, SmartSSD, and Newport CSD, have the capability to perform parallel computation. SmartSSD is an FPGA-based CSD and employs multiple computation units (CUs), up to 15 CUs. A CU is a computing instance created within the FGPA to execute the kernel.
Meanwhile, the Newport CSD is equipped with a quad-core ARM processor, enabling multi-kernel execution by multi-threading on Linux OS.
We first present the evaluation results of SmartSSD with an increasing number of CUs with selected big data workloads, and then we discuss the performance of Newport CSD in detail.

\begin{figure}[!t]
\centering
	\subfloat[SmartSSD]{
       \includegraphics[width=0.35\linewidth]{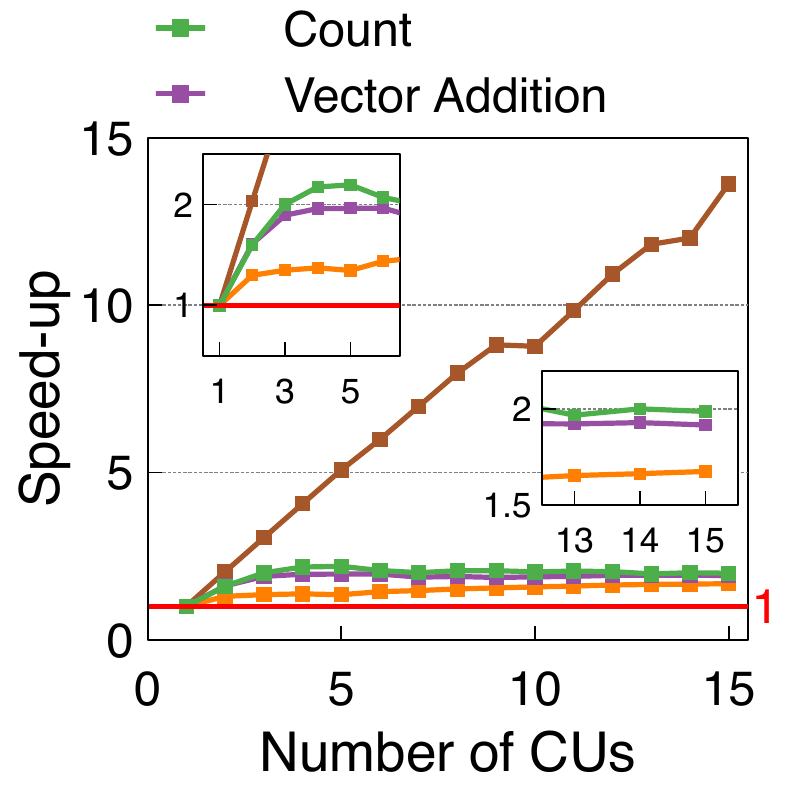}}
	\hfil
	\subfloat[Newport CSD]{%
       \includegraphics[width=0.35\linewidth]{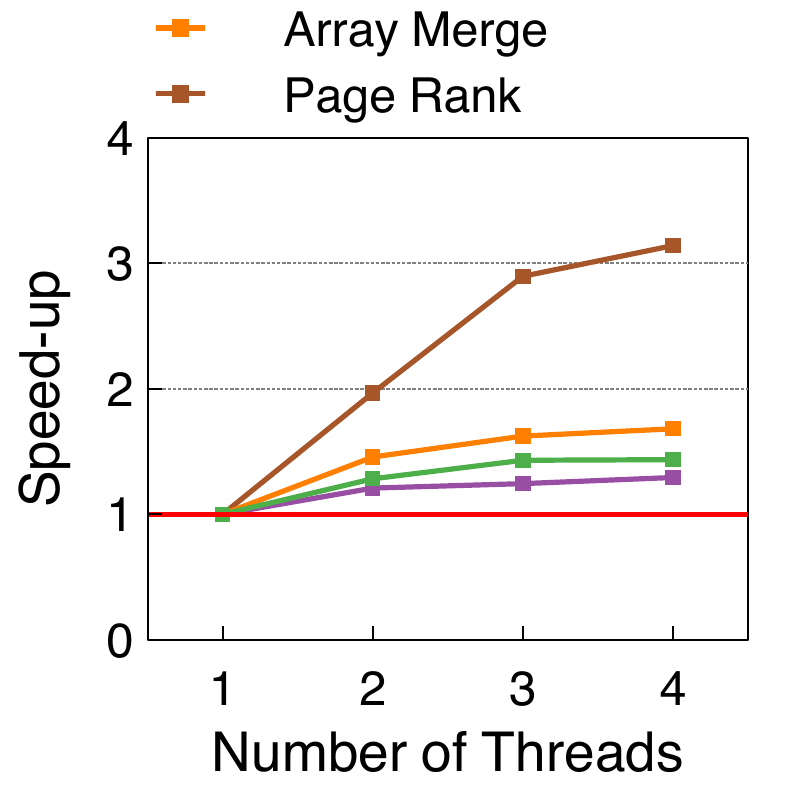}}
    \hfill
	\caption{
	{Impact of parallel computation by either multiple CUs used or multi-threading. 
	All experiments were normalized to the throughput when using one CU or thread. 
	}
    }
    \vspace{-5pt}
	\label{plot:csd_multi}
\end{figure}

{\bf SmartSSD's Results:} Figure~\ref{plot:csd_multi}(a) shows the throughput improvement (Speed-up) of SmartSSD with an increasing number of CUs (up to 15 CUs in our case). 
To fully exploit the performance characteristics of SmartSSD's FPGAs, we applied several performance optimization techniques, such as local memory buffers, loop pipelining, and array partitioning, to evaluate big analysis kernels~\cite{vitis3,vitis2,vitis,vitis_platform,vitis_doc}.
In Figure~\ref{plot:csd_multi}(a), the throughput of the Count and Vector Addition kernels improve up to 5 CUs by $2.2\times$ and $2.0\times$ respectively, after which the throughput is saturated.
The main reason for saturating throughput is that the CU executes the kernel after the data loaded in DRAM is copied to BRAM (Block RAM), and the memory copy from DRAM to BRAM becomes the bottleneck. 
BRAM~\cite{bram} is an FPGA's on-chip RAM, which can process data loaded in local memory faster than DRAM. Therefore, using BRAM is more effective than using DRAM at the expense of memory copy overhead. However, when many CUs access BRAM at the same time, BRAM access becomes a bottleneck.
On the other hand, Page Rank is a CPU-intensive workload, and there is almost no bottleneck caused by the memory copy mentioned above. Thus, the Page Rank shows perfect scaling as the number of CUs increases, improving up to $13.6\times$. 
Array Merge shows a gradual increase, but only up to 1.6$\times$. Like Page Rank, Array Merge is a CPU-intensive workload, but the algorithm has a lot of if-else statements for merge operations, which is the main impediment to performance gains.

{\bf Newport CSD's Results:} 
Figure~\ref{plot:csd_multi}(b) shows the evaluation results of the Newport CSD with multithreading enabled within analysis kernels.
Since Newport CSD is equipped with 4 CPU cores, we conducted experiments with up to 4 threads by mapping each execution thread to each core (one-to-one mapping).
All workloads used in each evaluation were written in parallel programs to enable multithreading. 
The results in Figure~\ref{plot:csd_multi}(b) are similar to those in Figure~\ref{plot:csd_multi}(a).
As shown in Figure~\ref{plot:csd_multi}(b), throughput scales up to 4 threads for all workloads.
The throughput of Count, Vector Addition, Array Merge, and Page Rank has been improved up to 1.4$\times$, 1.3$\times$, 1.7$\times$, and 3.1$\times$, respectively.
However, a notable observation is that Page Rank linearly scaled in SmartSSD (as shown in Figure~\ref{plot:csd_multi}(a)), but it only linearly scaled up to 3 threads in Newport CSD (as shown in Figure~\ref{plot:csd_multi}(b)), and thereafter, throughput improvement is slightly reduced (Figure~\ref{plot:csd_multi}(b)). 

Furthermore, in SmartSSD, throughput is improved in the order of Page Rank, Count, Vector Addition, and Array Merge. In contrast, in Newport CSD, throughput is improved by Page Rank, Array Merge, Count, and Vector Addition, showing different results. The reason is that SmartSSD's FPGA needs to apply various optimization techniques to achieve optimal throughput. The kernel code of an FPGA is more complex that is different from the code that runs on a typical ASIC-based processor. Therefore, the trend of throughput improvement may vary between kernels/workloads.

\subsubsection{Workload Classification}
\begin{figure}[!t]
	\centering
    \includegraphics[width=0.7\linewidth]{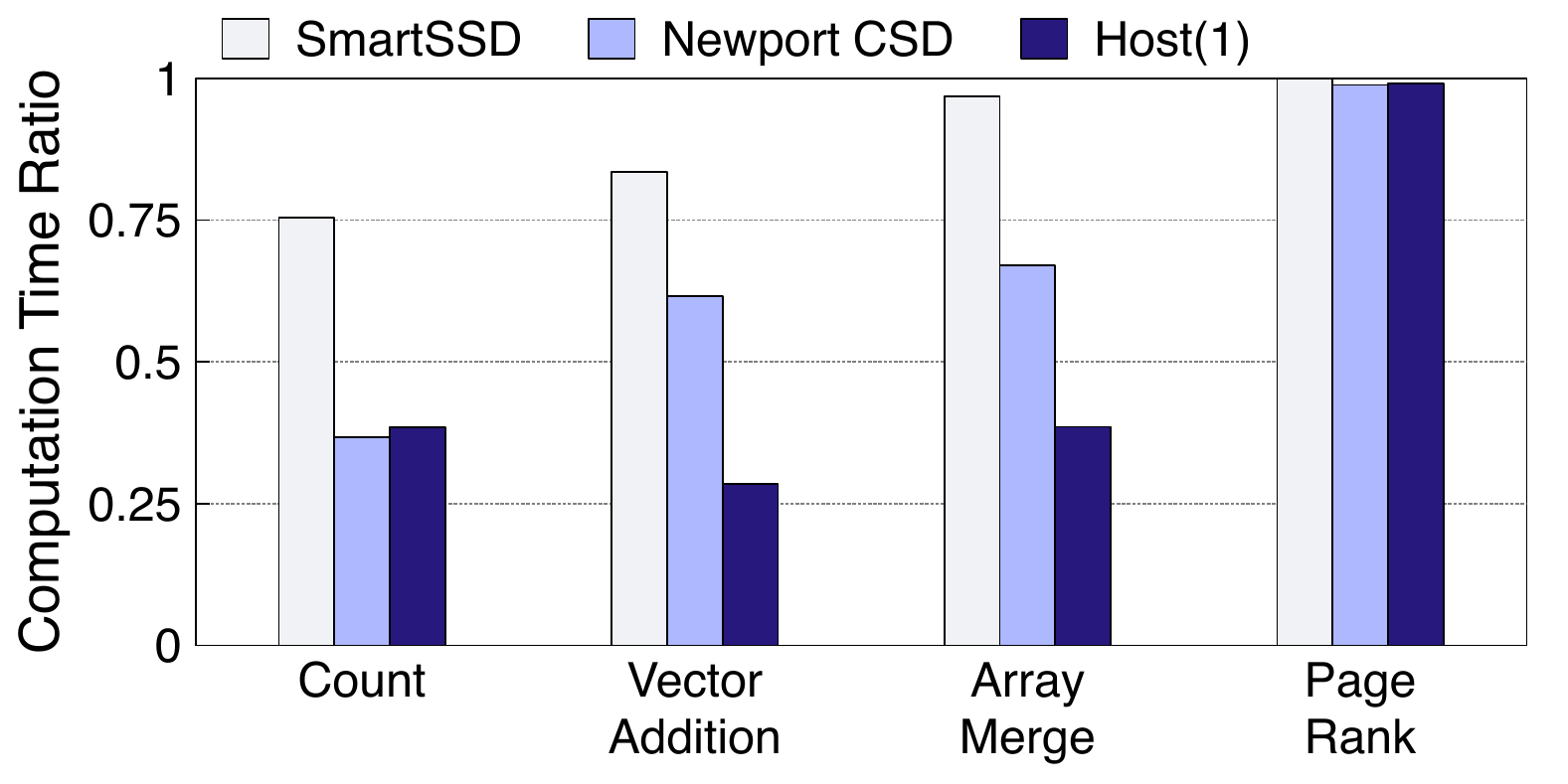}
	\caption{
    {Computation time ratio of execution time for each workload of SmartSSD, Newport CSD and host CPU(1)}. 
	}
    \vspace{-5pt}
	\label{plot:comp_ratio}
\end{figure}

As observed in Figure~\ref{plot:csd_multi}, the more parallel processing of the workload computation, the higher the throughput. However, as explained in Amdahl's law~\cite{Gustafson2011}, the throughput improvement has a limit bound to the data transfer time that cannot be further reduced. Therefore, we analyze computation and I/O ratios for each workload. Figure~\ref{plot:comp_ratio} shows the computation time ratio to total execution time (CTR) for each workload.
CTR values vary depending on the system. For the convenience of explanation, based on the CTR value of the host system, we classify workloads with a CTR of less than 0.5 as I/O-intensive workloads and otherwise as compute-intensive workloads. For example, Page Rank is a completely compute-intensive workload.

\subsubsection{CSD vs. Host System}

\begin{figure}[!t]
	\centering
    \includegraphics[width=0.7\linewidth]{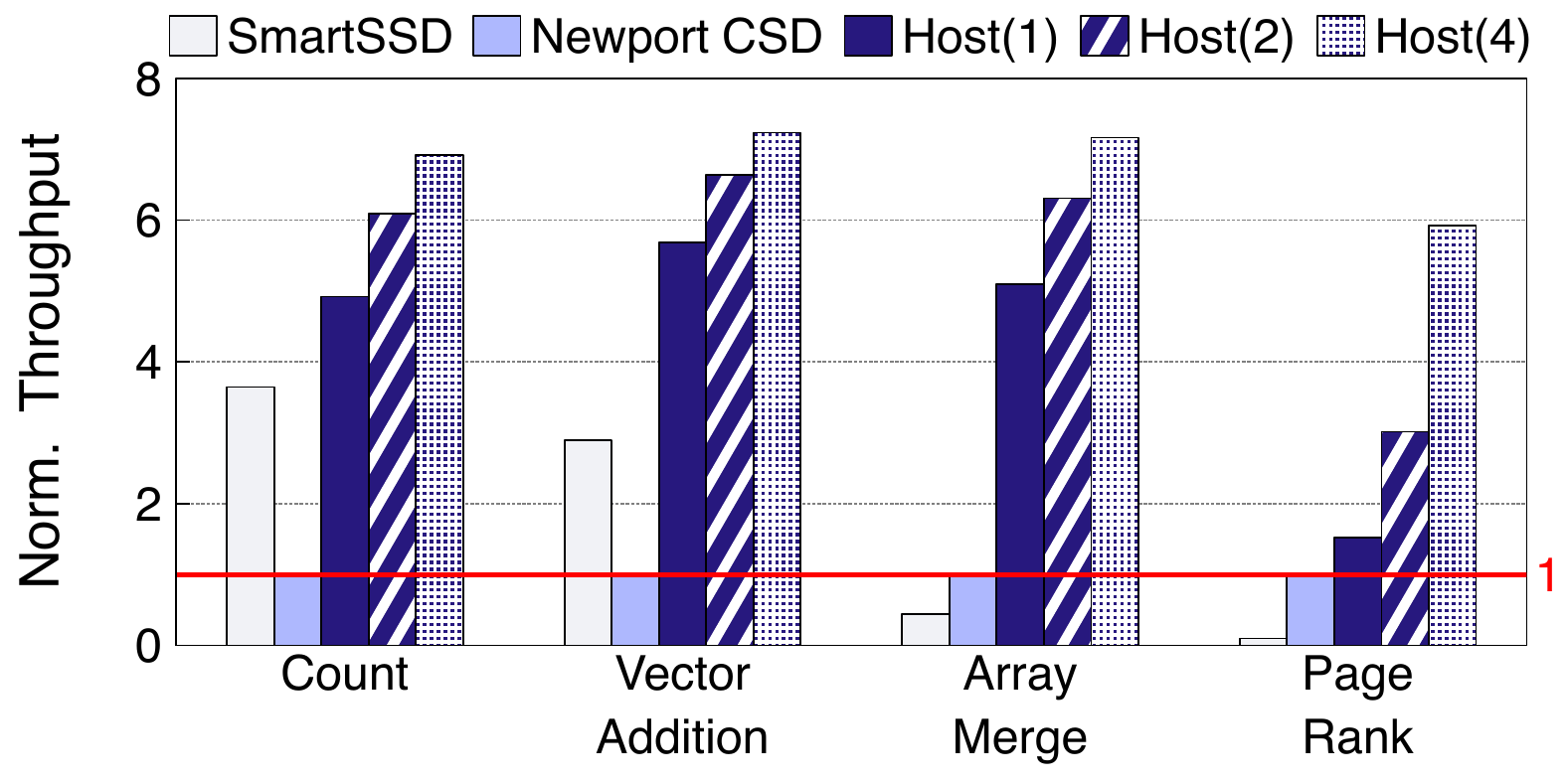}
	\caption{
    {Throughput comparison for each workload on SmartSSD, Newport CSD, and host systems. Throughput is normalized to the throughput of Newport CSD.
    In Host($n$), $n$ represents the number of active cores on the host system.} 
  	}
    \vspace{-5pt}
	\label{plot:comp_workload}
\end{figure}

Now we compare the throughput of a single CSD with the host system\footnote{Hereafter, we use the host system and SSD system interchangeably.}.
Figure~\ref{plot:comp_workload} shows the evaluation results for all workloads. Throughput was normalized to Newport CSD.
Each CSD is configured to achieve maximum throughput using multiple computational units (refer to Figure~\ref{plot:csd_multi}).
The host system used a SmartSSD as a block device.
We limit the number of cores for the host system to 4, and each configuration is represented as Host(1), Host(2), and Host(4) in Figure~\ref{plot:comp_workload}.

In Count and Vector Addition, 
SmartSSD has about 3$\times$ higher throughput than Newport CSD.
This is because both Count and Vector Addition are I/O intensive workloads, and although Newport CSD's computation latency is lower than SmartSSD's, throughput is more affected by internal I/O bandwidth (SmartSSD's internal I/O BW is higher, refer to Figure~\ref{plot:csd_comp_bw}).
In addition, Host(1) has a higher throughput of about 5$\times$ than Newport CSD. 
When compared to SmartSSD and Host(1), Host(1) has up to 1.3$\times$ and 2$\times$ higher throughput than SmartSSD.
On the other hand, as the number of active cores increases in the host system, the workload throughput improves by about 20\% per core. This is because it is bound to the data movement time between the host and the SSD.

In Array Merge and Page Rank, unlike observed in Count and Vector Addition, Newport CSD shows 2.2$\times$ and 10$\times$ higher throughput than SmartSSD because Newport CSD has higher computational power than SmartSSD. 
Although Array Merge is an I/O-intensive workload, it requires sufficient computational power as well. Thus, we can categorize the Array Merge kernel as compute- and I/O-intensive workload.
Also, as expected, the workload throughput of the host system is higher than that of CSD. However, it is noteworthy that the throughput of the host system is observed to be significantly higher in Page Rank. 
This is because Page Rank is a completely compute-intensive workload, and computational performance has the greatest impact on throughput. 
Therefore, the host system, which has the highest computational power, exhibits a significantly higher workload throughput than any CSD.

\begin{figure*}[t]
\centering
    \centering
    \begin{tabular}{cccc}
        \includegraphics[width=0.23\linewidth]{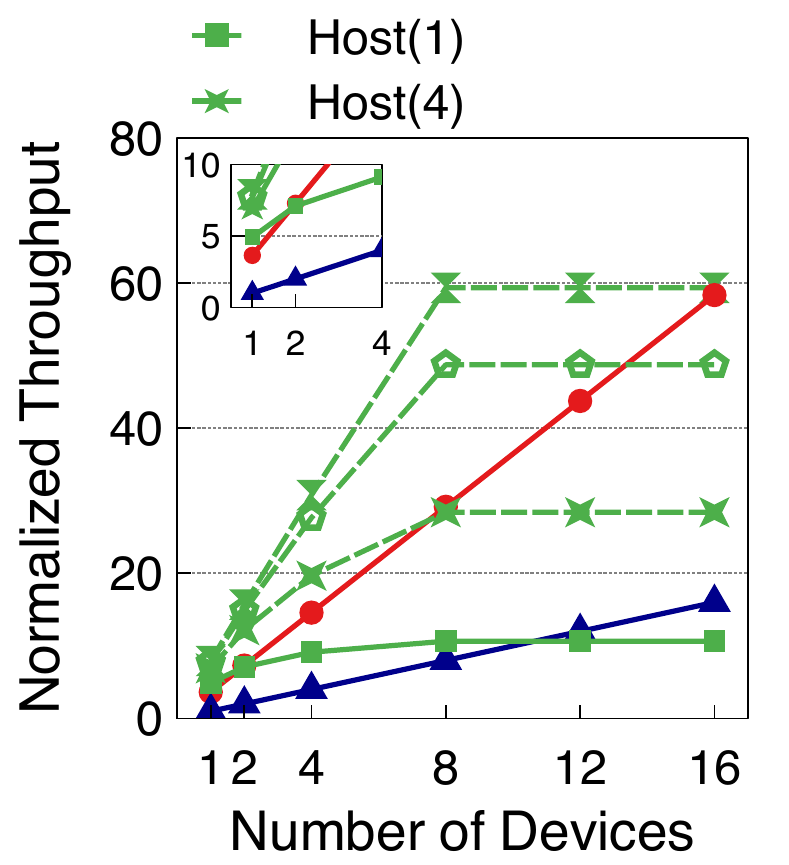} &
        \includegraphics[width=0.23\linewidth]{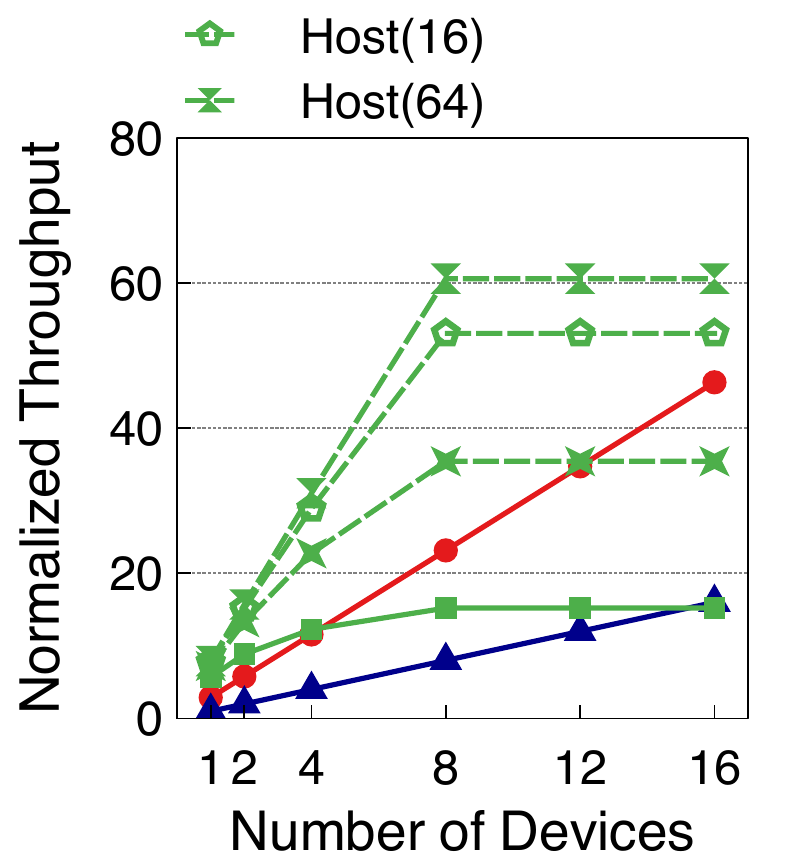} &
        \includegraphics[width=0.23\linewidth]{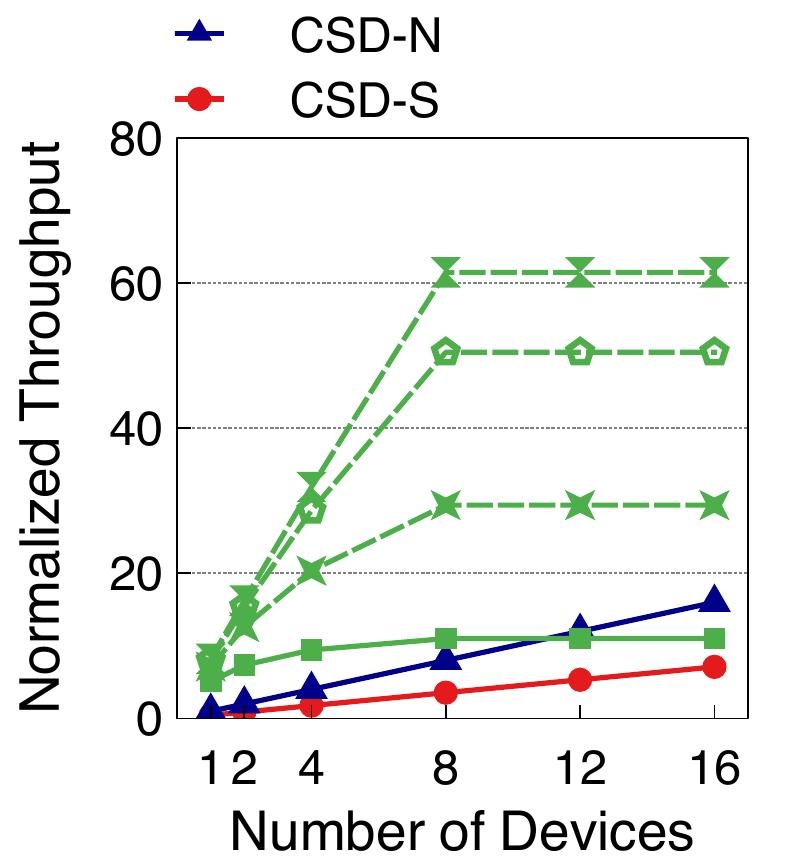} &
        \includegraphics[width=0.23\linewidth]{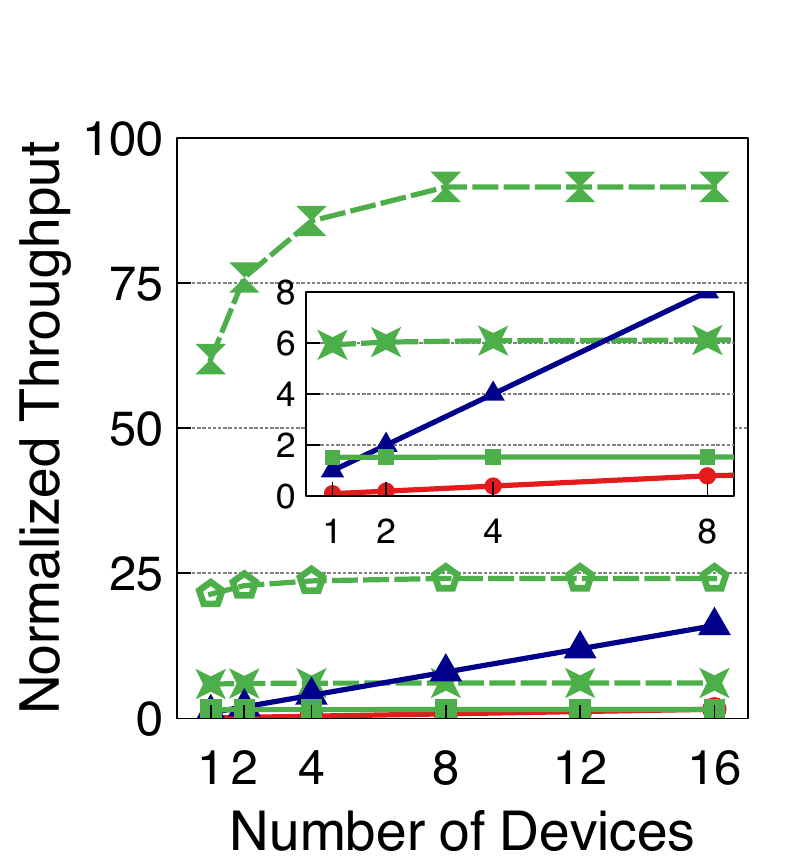}\\
        (a) Count &
        (b) Vector Addition &
        (c) Array Merge &
        (d) Page Rank \\
    \end{tabular}
	\caption{
	Evaluation to find the break-even point for normal conditions.
	In all results, throughput is normalized to the throughput of a single Newport CSD system for each workload. {In Host($n$), $n$ represents the number of active cores on the Host.}
	CSD-N and CSD-S denote CSD-array system (Newport CSD) and CSD-array system (SmartSSD), respectively.
    }
    \vspace{-10pt}
	\label{plot:model_normal_valid}
\end{figure*}

\subsection{Evaluating \cplan{} Solver}
\label{sec:eval_solver}

{\bf Finding the Break-Even Point (BEP):}
\cplan{} finds different BEPs depending on the workload, CSD-array, and host configurations.
To evaluate \cplan{}, we assumed that host's system supports PCIe~3.0 up to 32 lanes (1.0~GB/s $\times$ 32 = 32~GB/s), and each SSD uses 4~GB/s of bandwidth~\cite{spdk}. So, $k_\mathrm{limit}$ of Equation~(\ref{eq:host_array_limit}) is set to 8.

Figure~\ref{plot:model_normal_valid} shows each workload's throughput for various host configurations (1, 4, 16, or 64 active cores) and CSD-array systems with SmartSSD or Newport CSD. Throughput was normalized to a single Newport CSD.
For all workloads, CSD-array systems increase throughput linearly as the number of devices increases.
The host, on the other hand, has higher throughput with an increasing number of cores and storage devices (shown in $x$-axis of Figure~\ref{plot:model_normal_valid}) except Page Rank, but the host's throughput improvement was limited to 8 SSDs due to $k_\mathrm{limit}$ being 8. 
Since Page Rank is a compute-intensive workload, an increase in I/O throughput with an increasing number of storage devices does not lead to an improvement in throughput (more details on Page Rank are provided below).

Before analyzing the results, for convenience of explanation, we assume that Host(`a', `b') is a host system with `a' cores and `b' devices, and CSD-N(`c') is a CSD-array system composed of `c' Newport CSDs, and CSD-S('c') is a CSD-array system composed of `c' SmartSSDs.
In Count, Host(1, 1) has higher throughput than CSD-S(1). However, when the number of storage devices is 2, Host(1, 2) and CSD-S(2), the throughput meets for both configurations. 
This is because CSD-S has higher internal bandwidth, which leads to higher throughput with an increasing number of devices.
On the other hand, in CSD-N, the increase in throughput is lower than CSD-S as the number of devices increases, but the throughput increases gradually. Thus, CSD-N meets Host(1) when the number of devices reaches 12. 
As expected, Host(1) faces a limit in throughput scalability due to the PCIe bandwidth bottleneck, and eventually, CSD-N becomes higher than Host(1). 
{In addition, the host is equipped with a high-performance CPU, as shown in Table~\ref{tbl:host}, and when the workload is running a single thread, the CPU is very under-utilized. Therefore, as the number of cores increases at the host side, the throughput increases and the BEP value also increases.}
Host(4) meets CSD-S(8), and Host(16) meets CSD-S(13).
However, the degree of improvement in throughput is reduced.
This is because the workload execution time is bound to the data transfer time.

Array Merge is a mix of compute and I/O intensive workloads. That is, Array Merge requires a system with high computational power as well as high I/O throughput. Therefore, as shown in Figure~\ref{plot:model_normal_valid}(c), in both CSD-N and CSD-S, the increase in throughput is not higher than that of the host system as the number of devices increases.
In Figure~\ref{plot:model_normal_valid}(c), Host(1) and CSD-N meet when the number of devices is 12. CSD-S shows lower throughput than CSD-N. This is because the computational power of SmartSSD is lower than that of Newport CSD.
If the number of cores in the host increases, the host and CSD-array do not meet in 16 devices (refer to Host(4), Host(16), and Host (64) in Figure~\ref{plot:model_normal_valid}(c)). In other words, for Array Merge, 16 or more CSDs are required for the CSD-array system to achieve higher throughput than the host system. 
In summary, for I/O-intensive workload (Count,  Vector Addition and Array Merge), the host system's performance is limited by the PCIe bandwidth, where CSD-array catches up with host.

Finally, Page Rank is a completely compute-intensive workload. 
As shown in Figure~\ref{plot:model_normal_valid}(d), in the host system, the throughput increases with the number of cores, whereas the number of devices has little effect.
However, in 64 cores, the throughput increases with the number of devices. 
This is because the computation time is so low that the data transfer time affects the throughput.
On the other hand, CSD-N and CSD-S increase the workload throughput as the number of devices increases. As mentioned earlier, BEP increases as the number of cores in the host increases.

\begin{figure}[!t]
	\centering
    \includegraphics[width=0.55
    \linewidth]{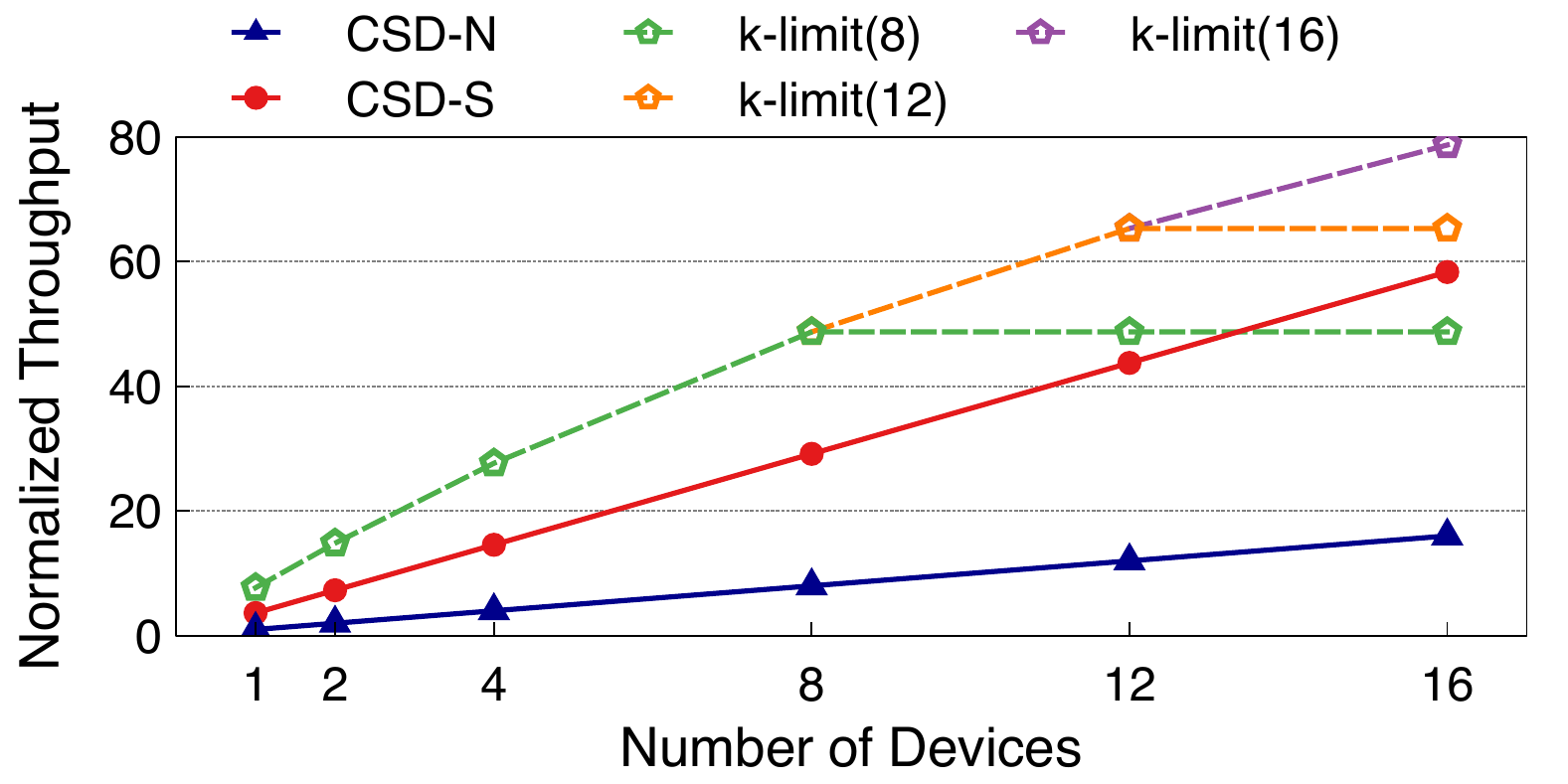}
	\caption{
    {{Example of change in maximum throughput of Host(16) according to $k_{limit}$ in Count.}} 
	}
    \vspace{-5pt}
	\label{plot:change_klimit}
\end{figure}

Figure~\ref{plot:change_klimit} shows an example of
the maximum throughput of Host(16) according to $k_{limit}$ of Equation~(\ref{eq:host_array_limit}) in Count. 
In the legend, 'n' in k-limit(n) means the $k_{limit}$ value.
The 'n' in k-limit(n) represents the value of $k_{limit}$.
{As the number of devices increases, the throughput of $k_{limit}$(8) improves to 8 devices, while the throughput of $k_{limit}$(12) and $k_{limit}$(16) improves to 12 and 16 devices, respectively.
Therefore, the BEP value also increases. The $k_{limit}$(8) meets CSD-S(13), while $k_{limit}$(12) and $k_{limit}$(16) meet after CSD-S(16).}
As such, \cplan{} can find BEP changes according to the number of devices that cause PCIe bottlenecks in various host systems.

\begin{figure*}[!t]
\centering
    \begin{tabular}{cccc}
       \includegraphics[width=0.23\linewidth]{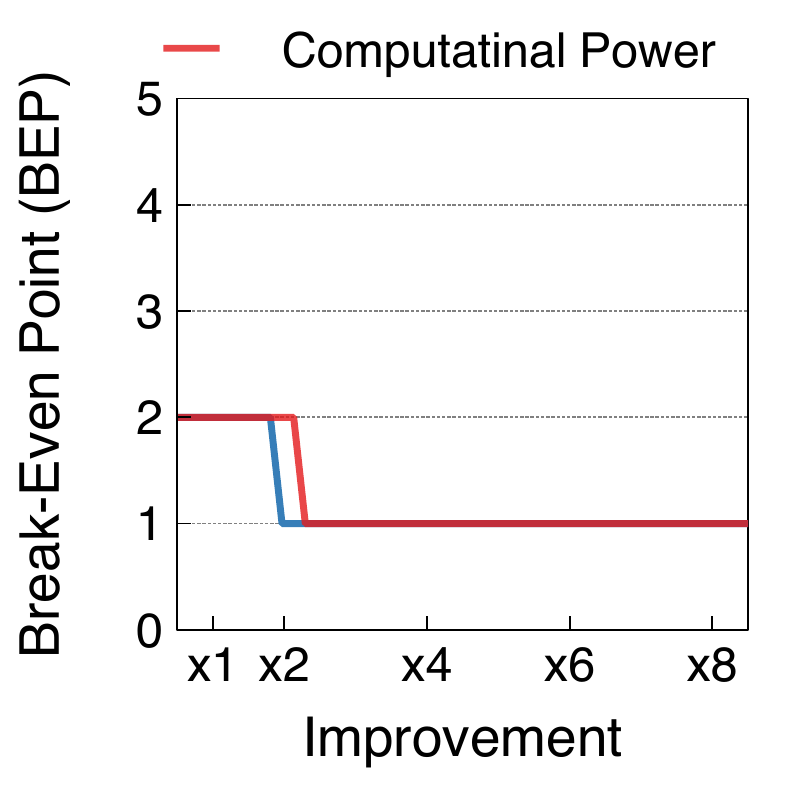} &
       \includegraphics[width=0.23\linewidth]{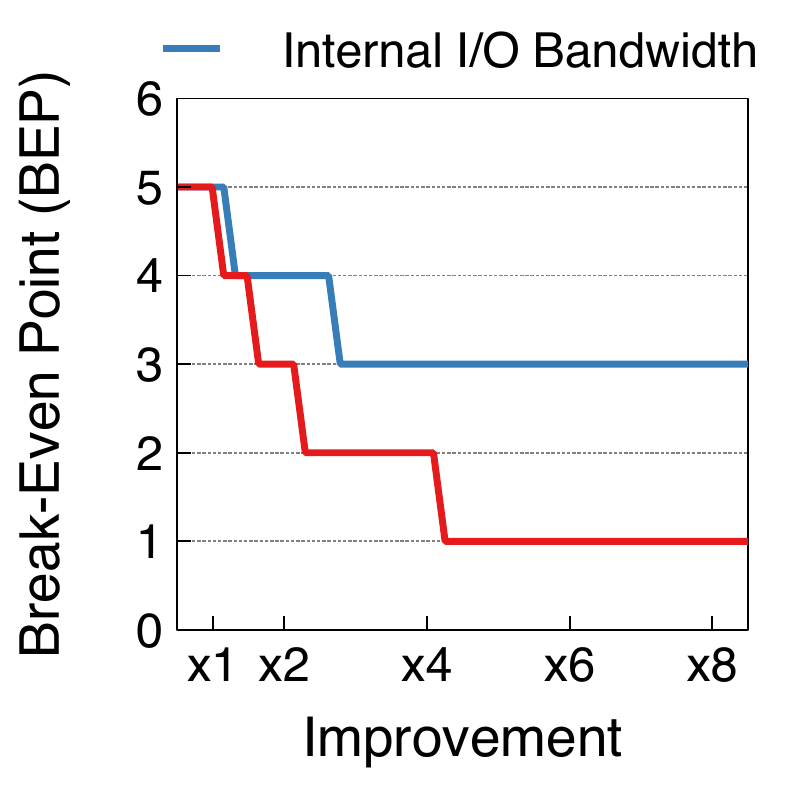} &       \includegraphics[width=0.23\linewidth]{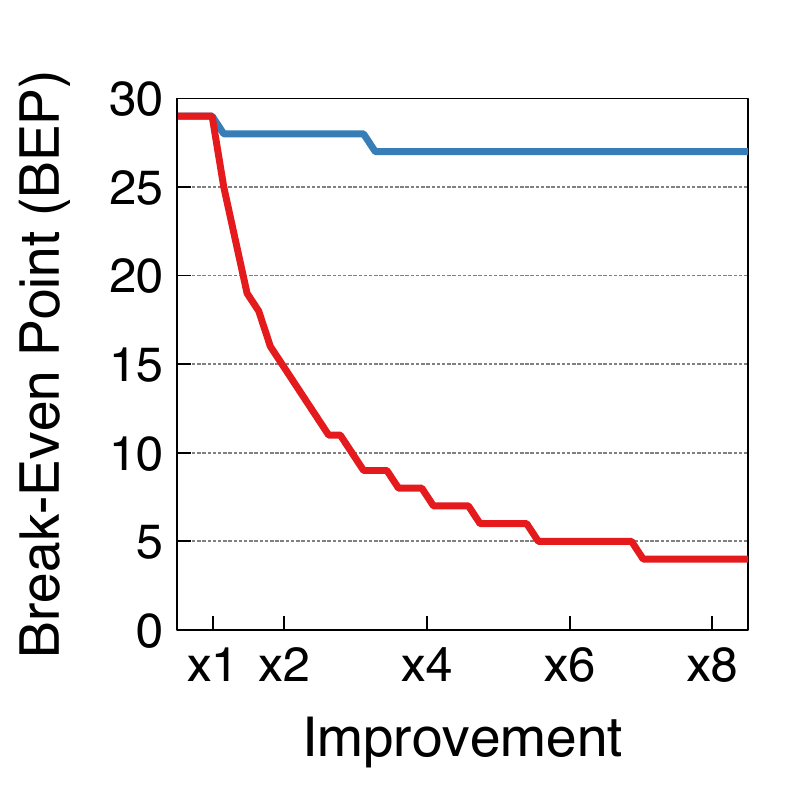} &
       \includegraphics[width=0.23\linewidth]{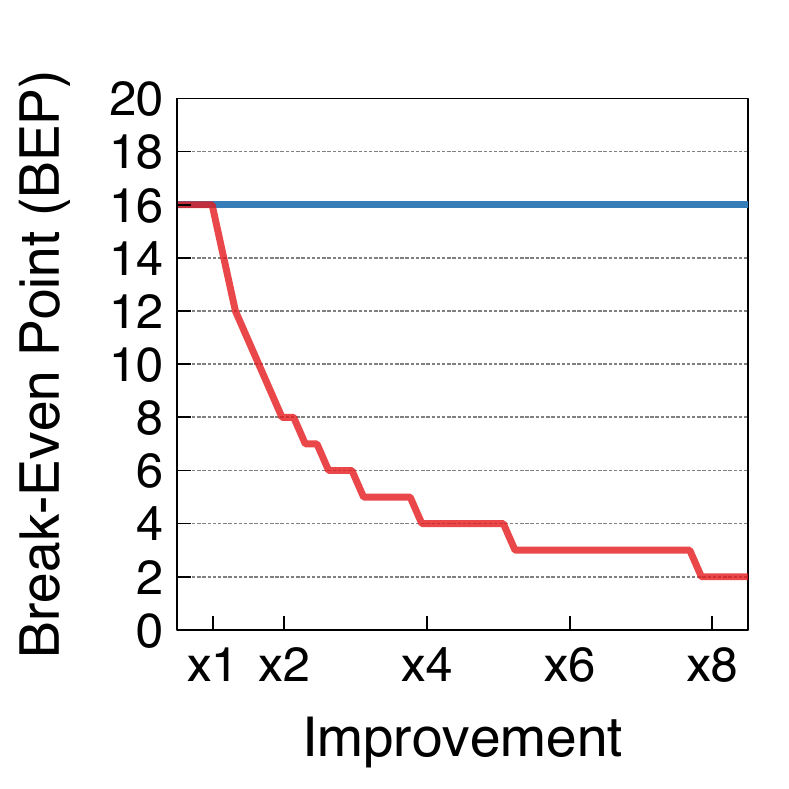}\\
       \vspace{-5pt}
       (a) Count & 
       (b) Vector Addition &
       (c) Array Merge &
       (d) Page Rank\\
       \multicolumn{4}{c}{SmartSSD-array system}\\
       \vspace{-5pt}
       \includegraphics[width=0.23\linewidth]{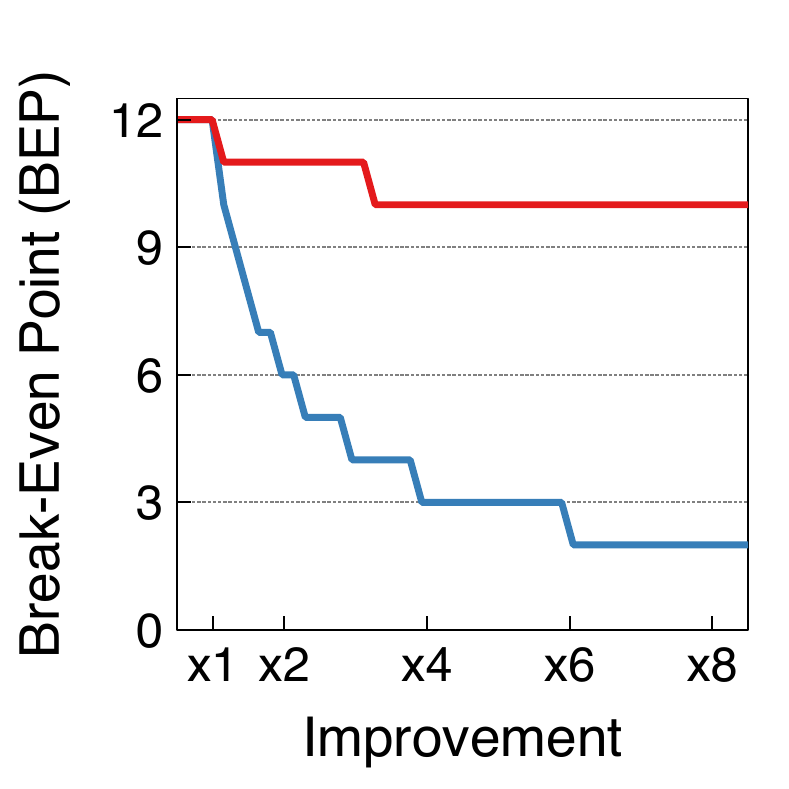} &
       \includegraphics[width=0.23\linewidth]{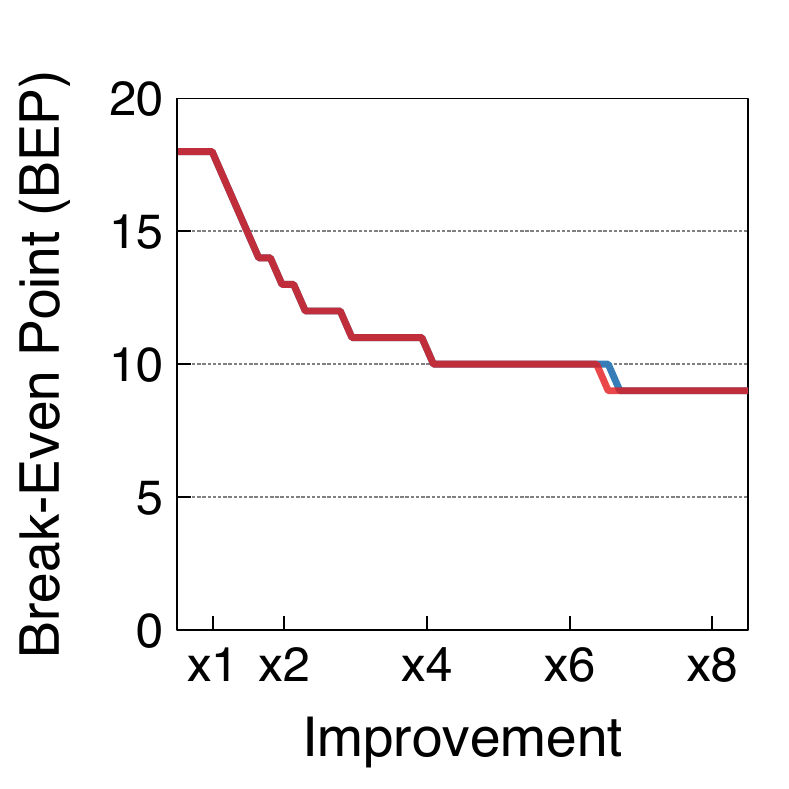} &
       \includegraphics[width=0.23\linewidth]{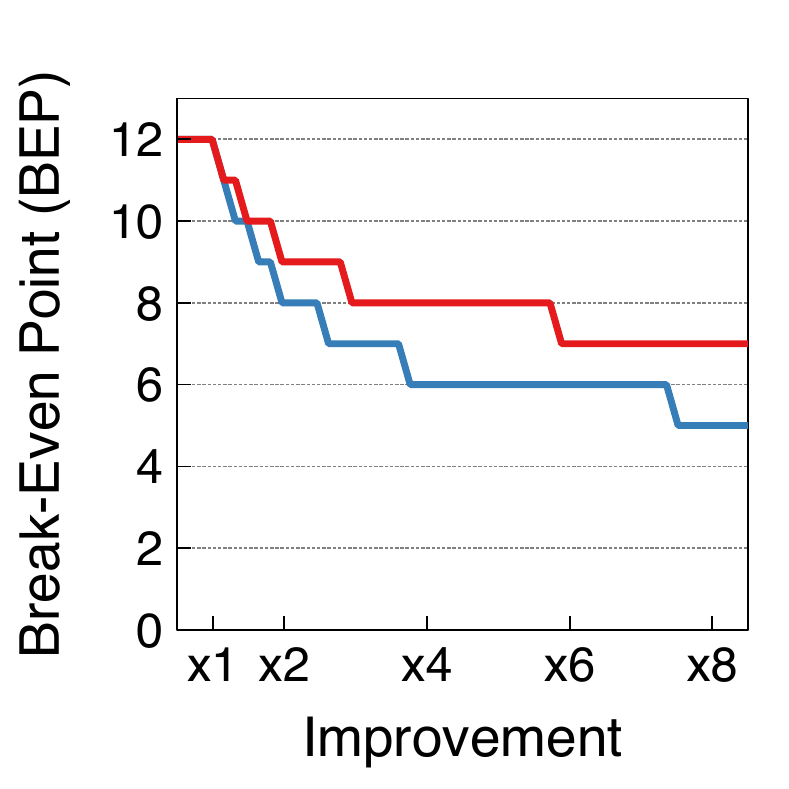} &
       \includegraphics[width=0.23\linewidth]{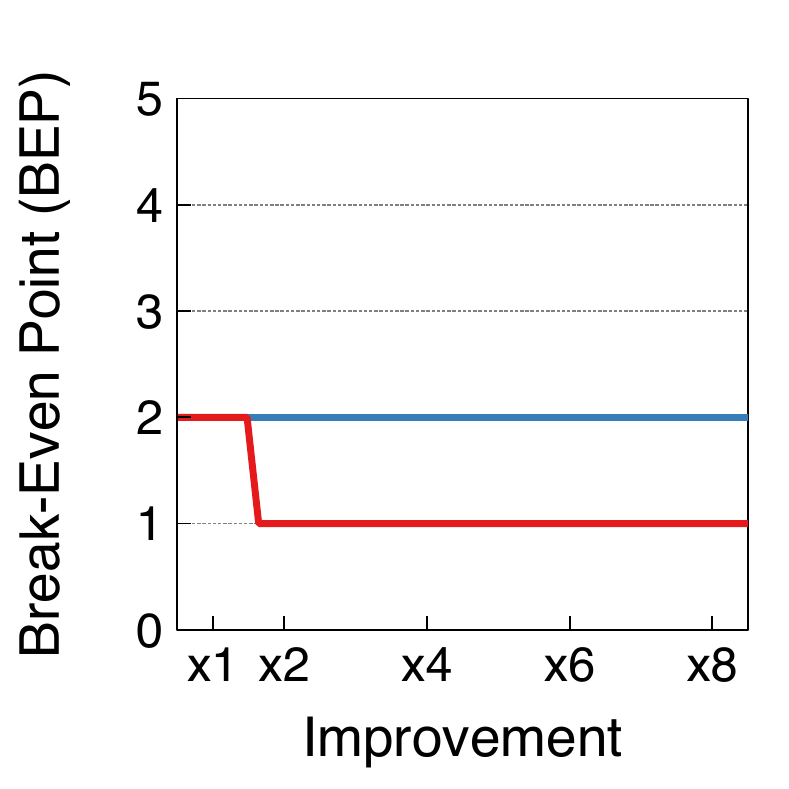}\\
       \vspace{-5pt}
       (e) Count & 
       (f) Vector Addition &
       (g) Array Merge &
       (h) Page Rank\\
       \multicolumn{4}{c}{Newport CSD-array system}\\
    \end{tabular}
	\caption{
	{
    	Analysis of reduction in the number of devices (BEP) by varying CSD's computational power or internal I/O bandwidth. 
    	}
    }
    \vspace{-10pt}
	\label{plot:model_proj_normal}
\end{figure*}

\noindent{\bf Impact of CSD Parameters:} The factors that determine the workload throughput in CSD are computational power and internal I/O bandwidth. \cplan{} can estimate the change in the BEP for Host($n$) according to the change in the values of these two factors.
In Equation~(\ref{eq:model_func_normal}), an increases in $R_\mathrm{tx}$ or $R_{(n)\mathrm{comp.}}$ means an increase in internal I/O bandwidth or computational power of CSD, respectively.
In this experiment, we assumed a Host($n$) system where $n$ is 1 (one active core on the server) and conducted experiments and analysis. 

Figure~\ref{plot:model_proj_normal}(a)-(d) shows the results for SmartSSD. SmartSSD has internal I/O bandwidth that meets the needs of each workload to some extent but has lower computational power. 
Therefore, increasing I/O bandwidth does not impact the BEP for compute-intensive workloads, such as Page Rank, while increasing the computational power does. 
On the other hand, for I/O intensive workloads, such as Count, SmartSSD shows sufficient internal I/O bandwidth and computational power. Thus changing any factor does not impact much in throughput. Vector Addition is also an I/O intensive workload, but increasing computational power does impact the BEP. Because SmartSSD has low computational power. 
Furthermore, Array Merge is a combination of compute and I/O intensive workloads. Thus, increasing both computational and I/O bandwidth will have a great effect on lowering the BEP.

\begin{figure}[!t]
	\centering
	\vspace{-10pt}
    \includegraphics[width=0.5\linewidth]{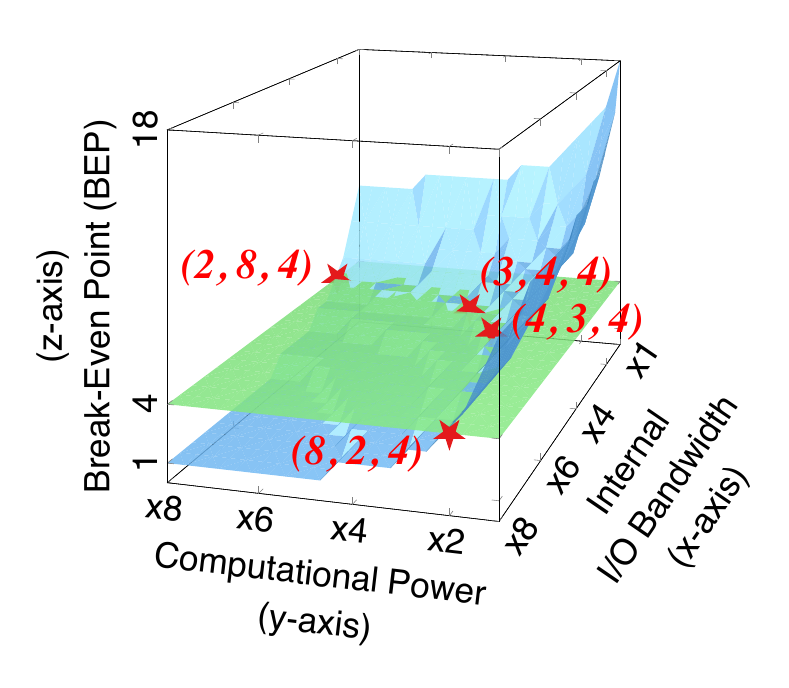}\\
	\caption{
    {Finding the values of BEP according to the change of internal I/O bandwidth and computation power of CSD for Vector Addition with the Newport CSD-array.
	}
	}
	\vspace{-5pt}
	\label{plot:model_normal_change_3d}
\end{figure}

Figure~\ref{plot:model_proj_normal}(e)-(h) shows the result for Newport CSD. Newport CSD has relatively higher computational power than SmartSSD but has lower internal I/O bandwidth. 
As shown in Figure~\ref{plot:model_proj_normal}, in I/O-intensive workloads such as Count, Vector Addition, and Array Merge, an increase in the internal I/O bandwidth has a great effect on lowering the BEP. However, in Vector Addition and Array Merge, computational power also tends to be affected. That is, Newport CSD has higher computational power than SmartSSD, but it is still lower than the host.
Finally, Page Rank, again, is a completely compute-intensive workload, so increasing I/O bandwidth has little effect on lowering BEP.

\cplan{} can find the BEP values that change according to the hardware parameters (computational power and internal I/O bandwidth) of the CSD. Figure~\ref{plot:model_normal_change_3d} visually shows the BEP values that \cplan{} finds. The $x$-axis means internal I/O bandwidth, the $y$-axis means computational power, and the $z$-axis means BEP (the number of devices) that \cplan{} finds. As shown in Figure~\ref{plot:model_normal_change_3d}, a 3D plane (drawn by a 3D function of $z=f(x,y)$) shows the changes in BEP values for $x$ and $y$ (blue plane). Also, the points where the green plane and the blue plane intersect in the figure are all combinations of hardware parameters corresponding to $\text{BEP}{=}C$ where $C$ is a constant. For example, in the Figure~\ref{plot:model_normal_change_3d}, $C$ is 4.
In the figure, four combinations that satisfy $\text{BEP}{=}4$\{(2, 7, 4), (3, 4, 4), (4, 3, 4), (8, 2, 4)\} are marked with red stars. 

Moreover, the results presented in Figure~\ref{plot:model_normal_change_3d} can be considered as the design guideline for CSD manufacturers. The two factors impacting the performance of CSD are: internal I/O bandwidth and computational power, and both are required to be improved.
CSD has very low computational power compared to the host CPU. In order to improve the performance of CSD, it is necessary to install a processor with higher computational power.

\subsection{\cplan{} Solver in Overload Situations}
\label{sec:impact_ol}
As mentioned in Section~\ref{sec:csdllan}, the host system can be overloaded due to excessive resource usage by applications co-located on the host.
In this section, we show how \cplan{} finds the BEP of a CSD-array-based system under such an overloaded host system.

First, as described in Section~\ref{sec:csd_perf}, the system architect should perform a performance characterization of the host system under overloaded conditions. An overload situation can occur for some reasons; lack of CPU cycles, insufficient memory, I/O bandwidth, or a mixture of these. In big data applications, overloaded situations often occur due to insufficient memory, 
thus, we consider it as the main cause here as well.
In this subsection, we provide guidelines to system architects for the experiments to be performed in overloaded conditions. 

The goal of the experiment is to find the optimal values of slow-down factors (mentioned in Equation~\ref{eq:model_prim}) and find BEP in various overloaded conditions.
{For this experiment, we simulate the overloaded condition on the host system by controlling the amount of physical memory by adopting $mlock()$\footnote{mlock() locks part or all of the calling process's virtual address space into RAM, preventing that memory from being paged to the swap area.} to limit the available memory to the analysis kernel. 
We simulated with 2~GB to 10~GB of available memory for analysis kernels. }

\begin{figure*}[t]
    \centering
    \begin{tabular}{cccc}
        \includegraphics[width=0.23\linewidth]{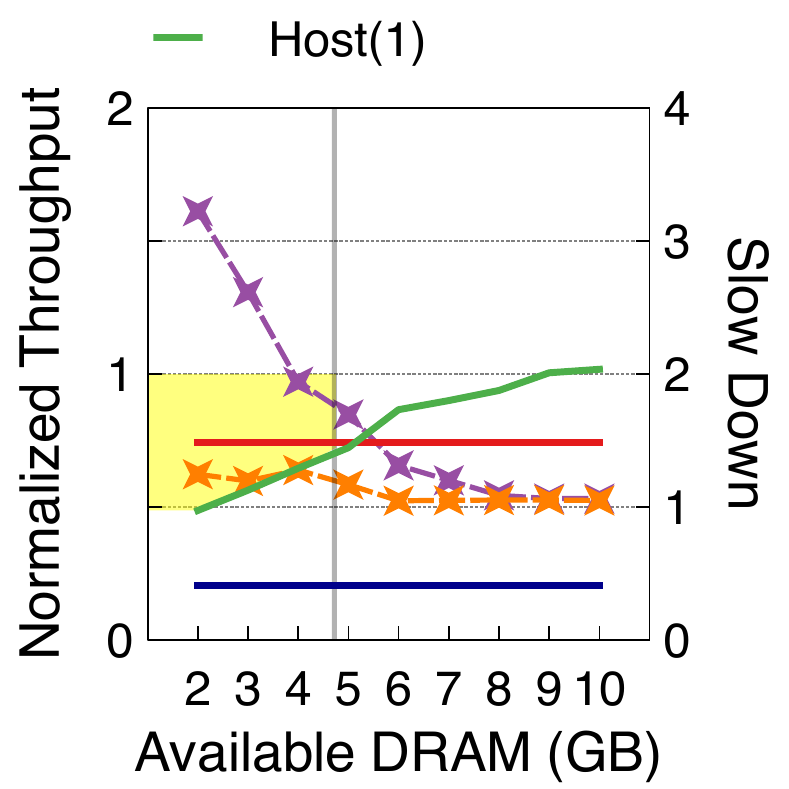} &
        \includegraphics[width=0.23\linewidth]{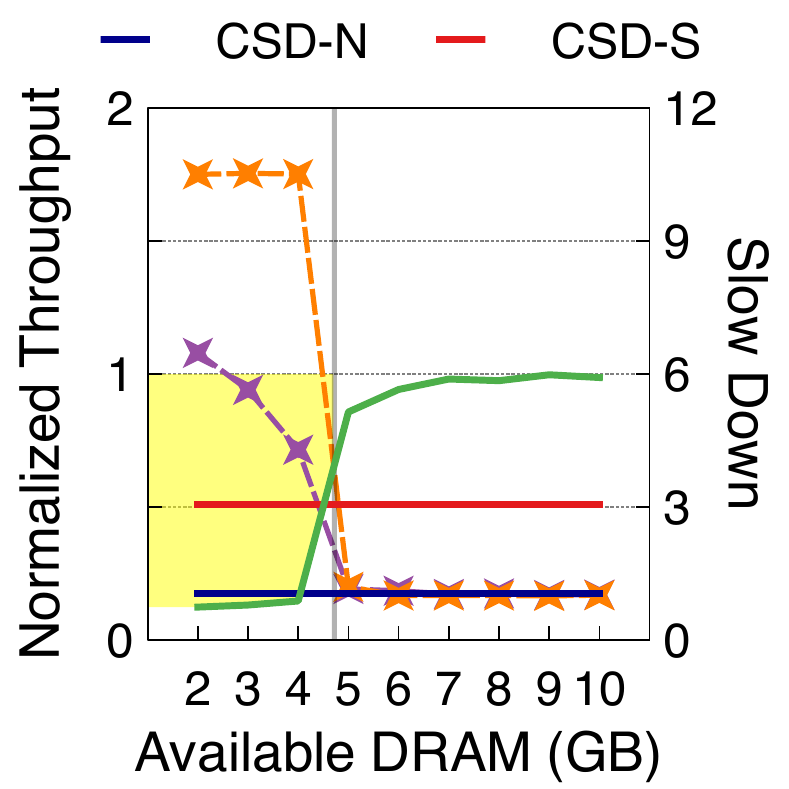} &
        \includegraphics[width=0.23\linewidth]{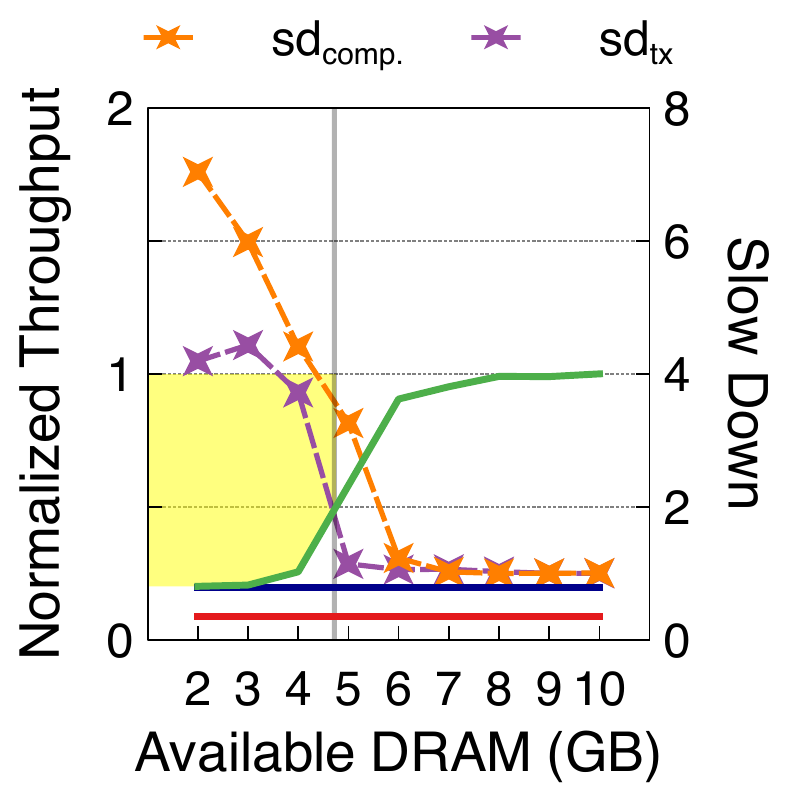} &
        \includegraphics[width=0.23\linewidth]{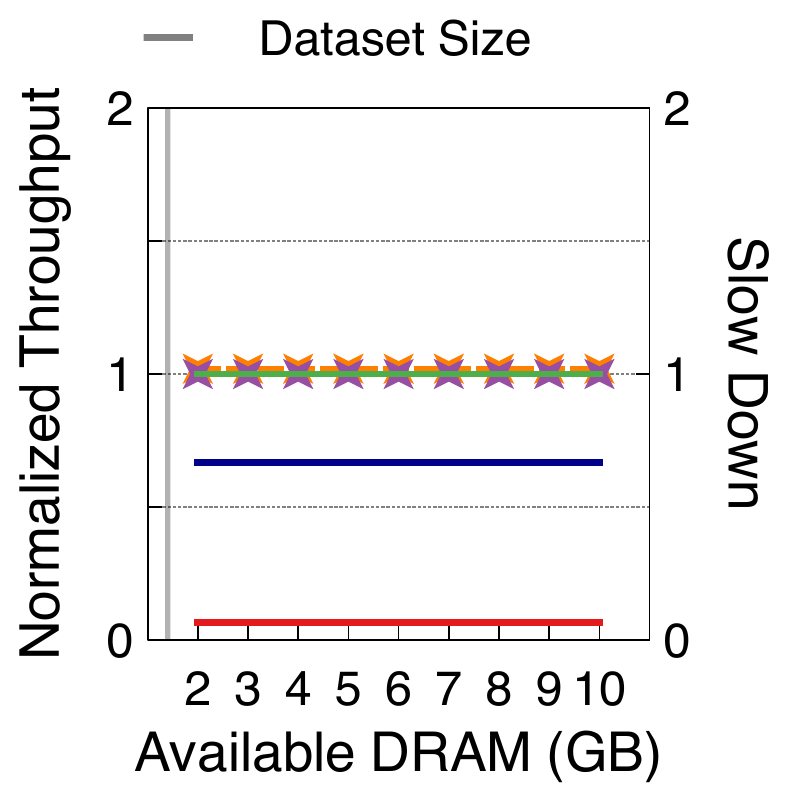}\\
        (a) Count &
        (b) Vector Addition &
        (c) Array Merge &
        (d) Page Rank \\
    \end{tabular}
	\caption{Comparison of the performance of analysis kernels for SSD system and CSD system under overload condition. In order to simulate the overload condition of the host machine, the amount of physical memory available to the analysis kernel is limited. 
	}
    \vspace{10pt}
	\label{plot:model_busy_throughput}
	    \centering
    \begin{tabular}{cccc}
        \includegraphics[width=0.23\linewidth]{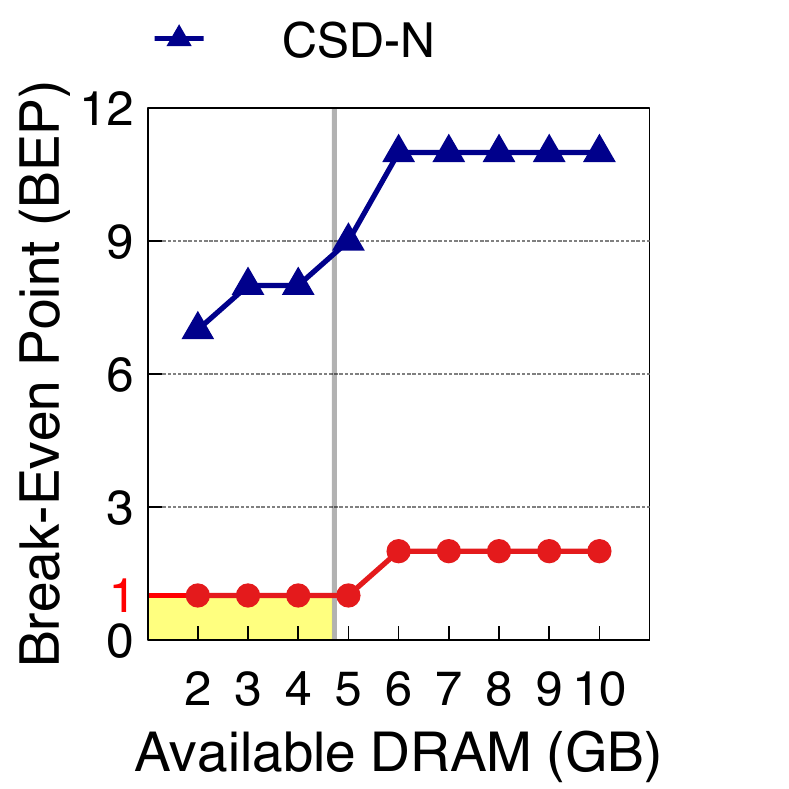} &
        \includegraphics[width=0.23\linewidth]{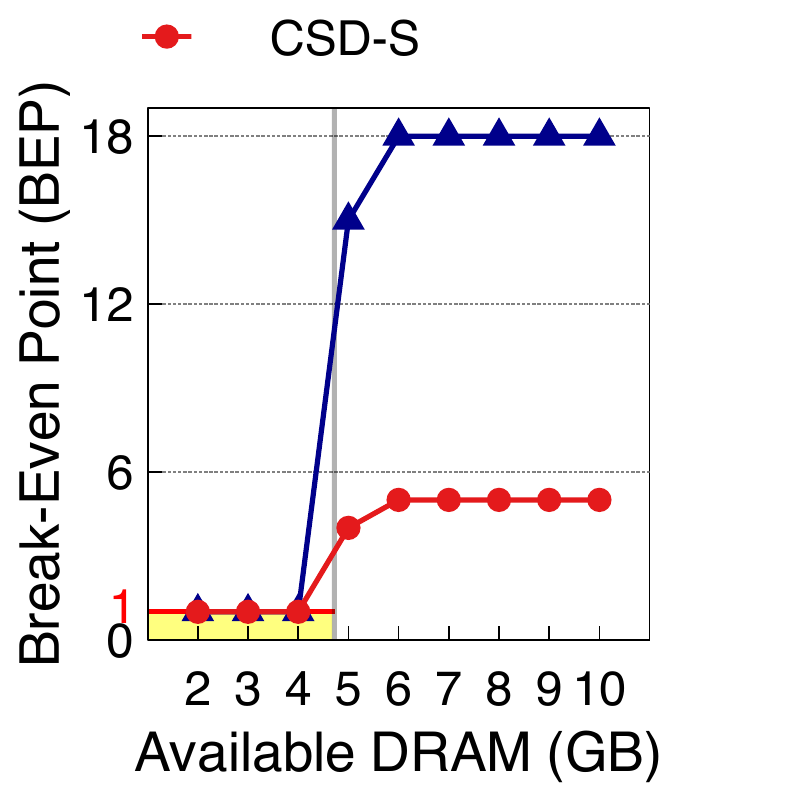} &
        \includegraphics[width=0.23\linewidth]{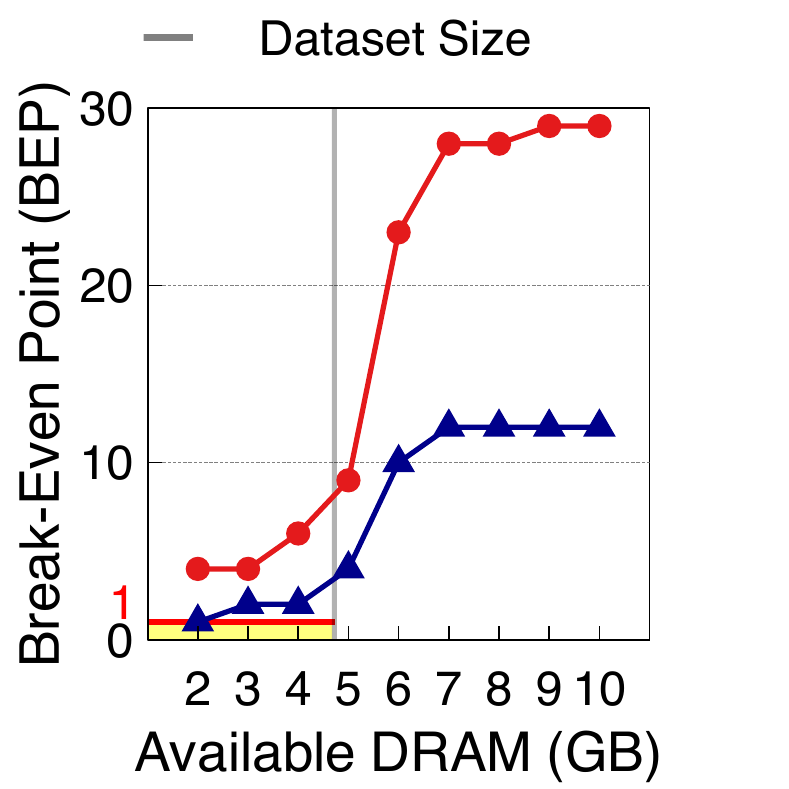} &
        \includegraphics[width=0.23\linewidth]{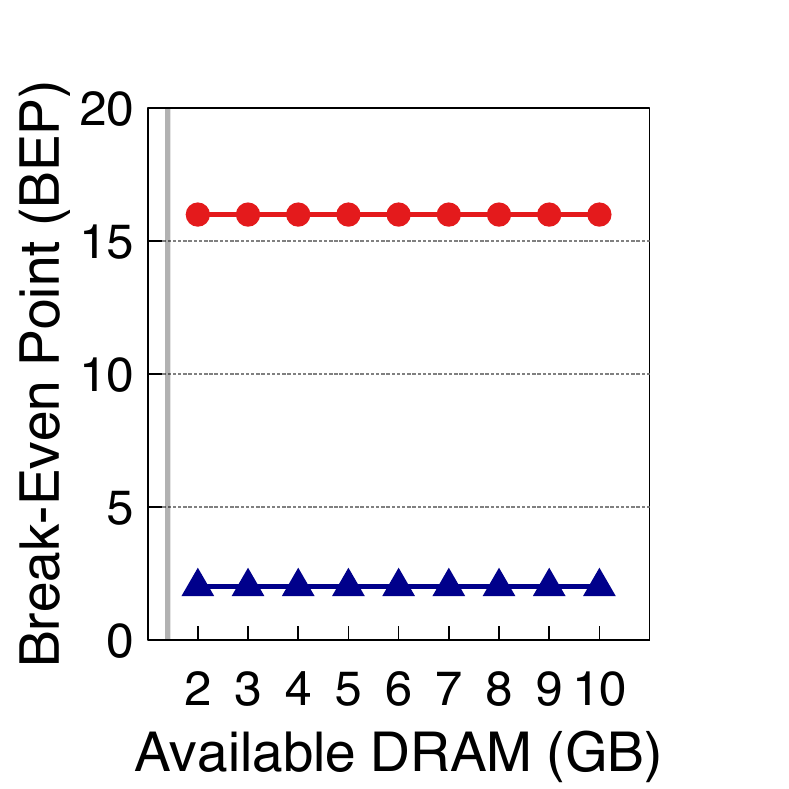}\\
        (a) Count &
        (b) Vector Addition &
        (c) Array Merge &
        (d) Page Rank \\
    \end{tabular}
	\caption{
	Results of finding deceleration factors and break-even points according to available DRAM size. 
    }
    \vspace{-10pt}
	\label{plot:model_busy_sd}
\end{figure*}

{\bf Results:} Figure~\ref{plot:model_busy_throughput} shows the comparison results of throughput according to the size of the host's available memory for the three systems {and shows the host's two slow-down factors ($sd_\mathrm{comp.}$, $sd_\mathrm{tx}$) from Equation~(\ref{eq:model_prim})}.
An increase in $sd_\mathrm{comp.}$ and $sd_\mathrm{tx}$ means the degradation of computational power and data transfer time of the host, respectively.
Here, we use the Host(1), CSD-N, and CSD-S notations used in Section~\ref{sec:eval_solver}.
All system's throughputs were normalized to the Host(1)'s throughput under normal conditions.
In Figure~\ref{plot:model_busy_throughput}(a)-(c), the Host(1) throughput is significantly reduced when the host's available memory is less than the dataset size, and host's slow-down factors are significantly increased.
{In Count, only $sd_\mathrm{tx}$ increases significantly because it is an I/O-intensive workload. Vector Addition and Array Merge are also I/O-intensive workloads, but $sd_\mathrm{comp.}$ and $sd_\mathrm{tx}$ grow together. This is because both have much computation compared to Count and is fatally affected when host's memory is insufficient.}
On the other hand, CSDs throughput are not affected by the availability of resources on the host.
In Figure~\ref{plot:model_busy_throughput}(d), the Host(1)'s throughput {and slow-down factors} does not change at all, no matter how much memory is available.
This is because the size of the dataset for the Page Rank workload is small. All datasets are loaded into memory, so disk swapping does not occur at all in the virtual memory system.

Figure~\ref{plot:model_busy_sd} shows the change of BEP according to the host's available memory for each workload. 
The result was calculated by substituting the slow-down factor used in Figure~\ref{plot:model_busy_throughput} into Equation~(\ref{eq:model_prim}).

In the Count, Vector Addition, and Array Merge {(except for Newport CSD's Count and SmartSSD's Array Merge)}, the BEP is 1 when the host's available memory is smaller than the dataset size and increases rapidly when the host's available memory begins to exceed the dataset.
This shows that the host's resource (memory) actually has a significant effect on the slow-down factor, and our proposed modeling from Equation~(\ref{eq:ssd_overload}) is well applied.

The parts marked in yellow in Figure~\ref{plot:model_busy_throughput} and \ref{plot:model_busy_sd} correspond to the case where the BEP is 1. 
In all cases, the BEP included in the parts marked in yellow in Figure~\ref{plot:model_busy_throughput} is equally included in Figure~\ref{plot:model_busy_sd}. 
This means that our proposed modeling works well. 
{Through this result, it is possible to analyze the effectiveness of CSD according to the change of host resources using \cplan{}.}

\noindent{\bf Impact of Host's Overloading:} 
\cplan{} can find the BEP value according to the change in the host's slow-down factors. For this, \cplan{} uses the following function:
\begin{equation}
\label{eq:model_func_overloaded}
\begin{aligned}
S_{\mathrm{overload}}(sd_\mathrm{tx}, \; sd_\mathrm{comp.}) = \text{max} \left( \left\lceil sd^{-1}_\mathrm{comp.} {\cdot} \{ (R^{-1}_\mathrm{tx} - sd_\mathrm{tx}) {\cdot} R_{(n)\mathrm{SSD}} {+} R^{-1}_{(n)\mathrm{comp.}}  ) \} \right\rceil ,1 \right) 
\end{aligned}
\end{equation}
The above function extends Equation~(\ref{eq:model_prim}) and has a form similar to Equation~(\ref{eq:model_func_normal}).
Here we show that \cplan{} uses the above function to find the value of BEP according to host's two slow-down factors.
For the experiment, we assumed a Host($n$) system where $n$ is 1.

\begin{figure*}[!t]
\centering
    \begin{tabular}{cccc}
       \includegraphics[width=0.23\linewidth]{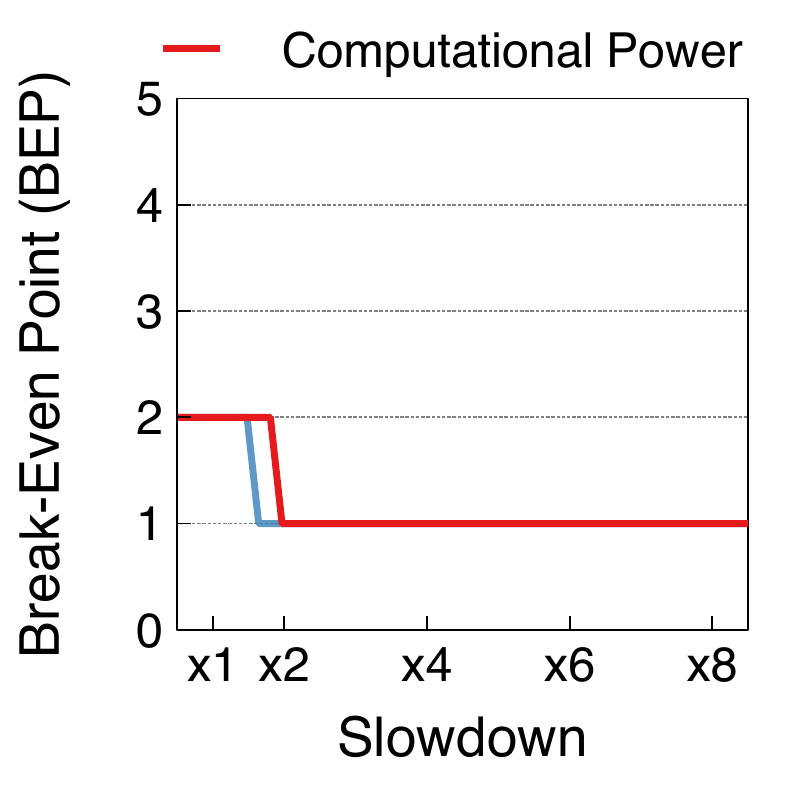} &
       \includegraphics[width=0.23\linewidth]{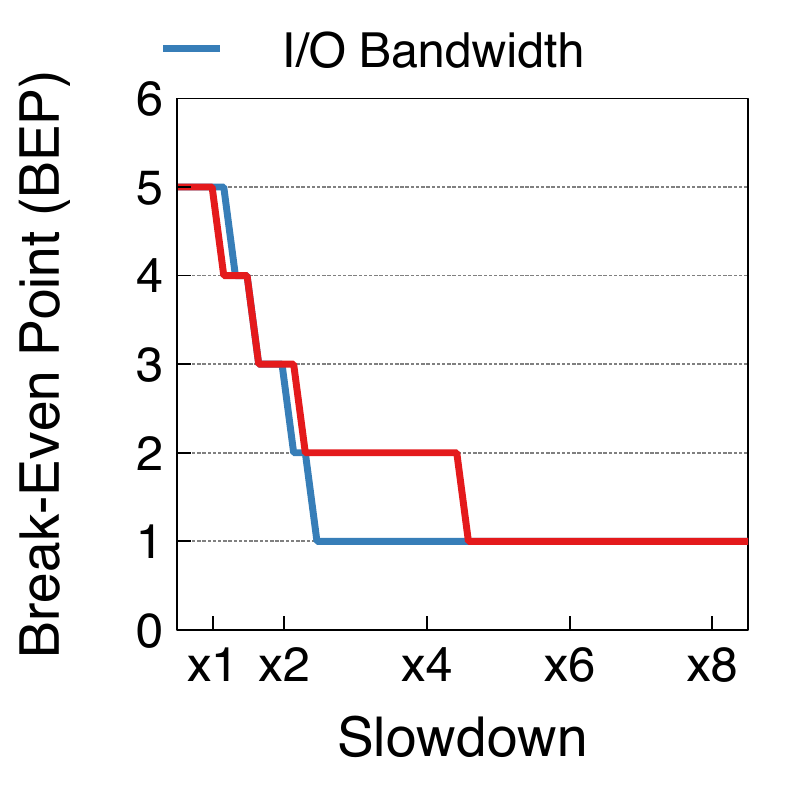} &       \includegraphics[width=0.23\linewidth]{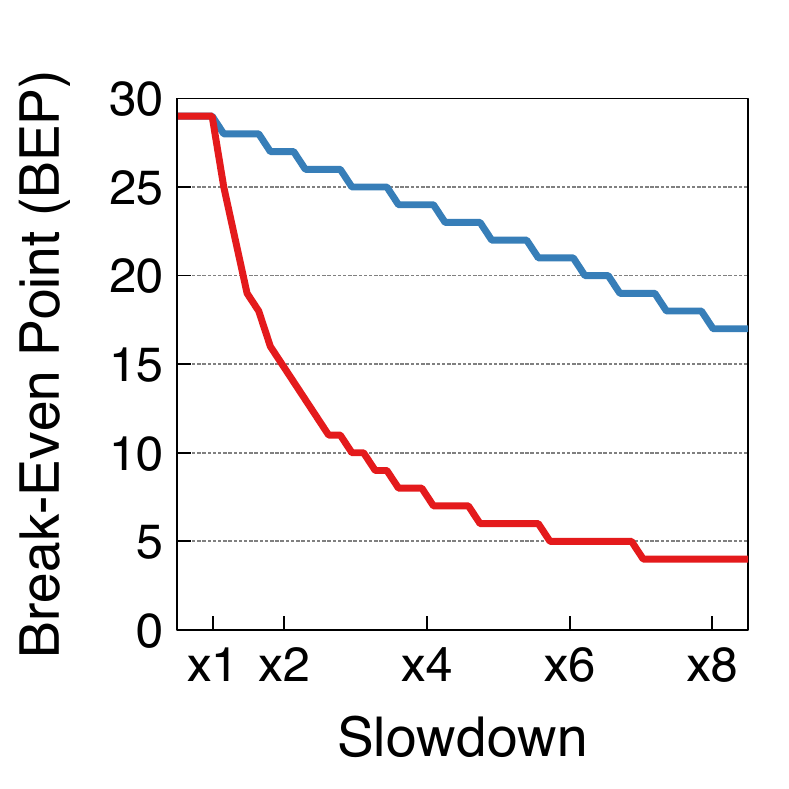} &
       \includegraphics[width=0.23\linewidth]{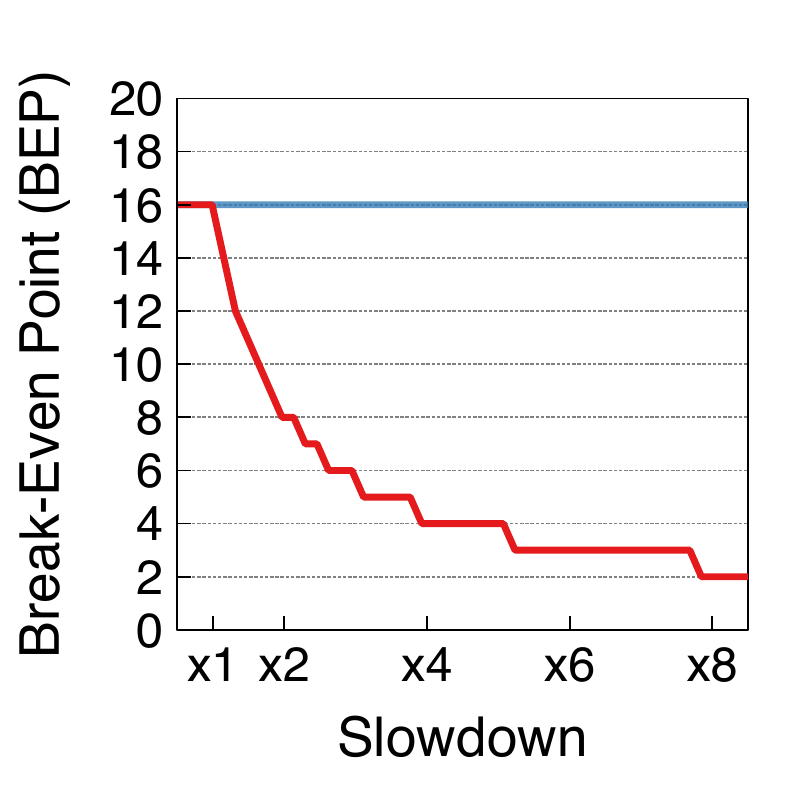}\\
       \vspace{-5pt}
       (a) Count & 
       (b) Vector Addition &
       (c) Array Merge &
       (d) Page Rank\\
       \multicolumn{4}{c}{SmartSSD-array system}\\
       \vspace{-5pt}
       \includegraphics[width=0.23\linewidth]{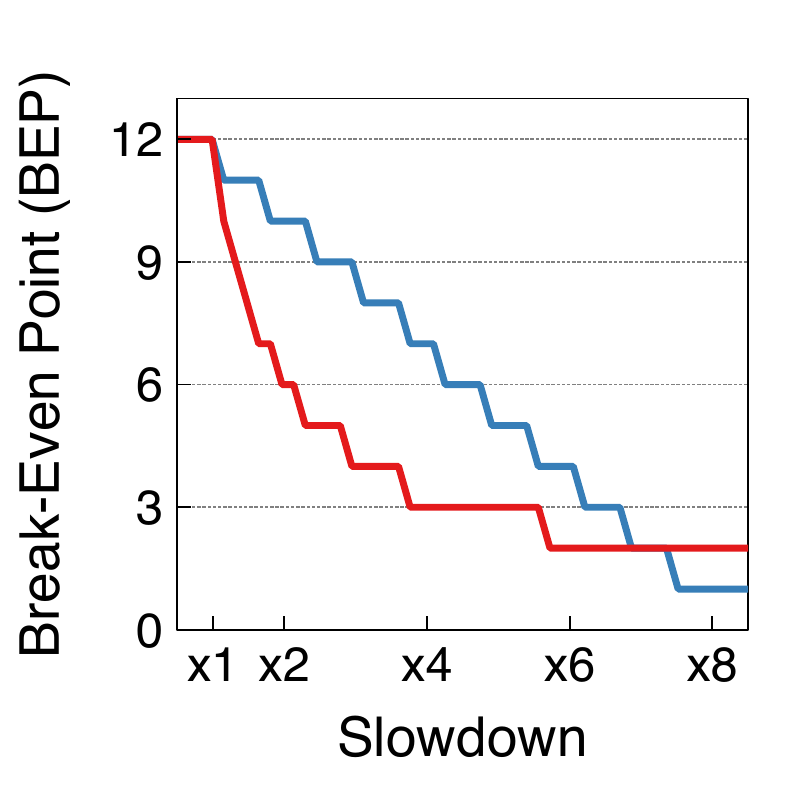} &
       \includegraphics[width=0.23\linewidth]{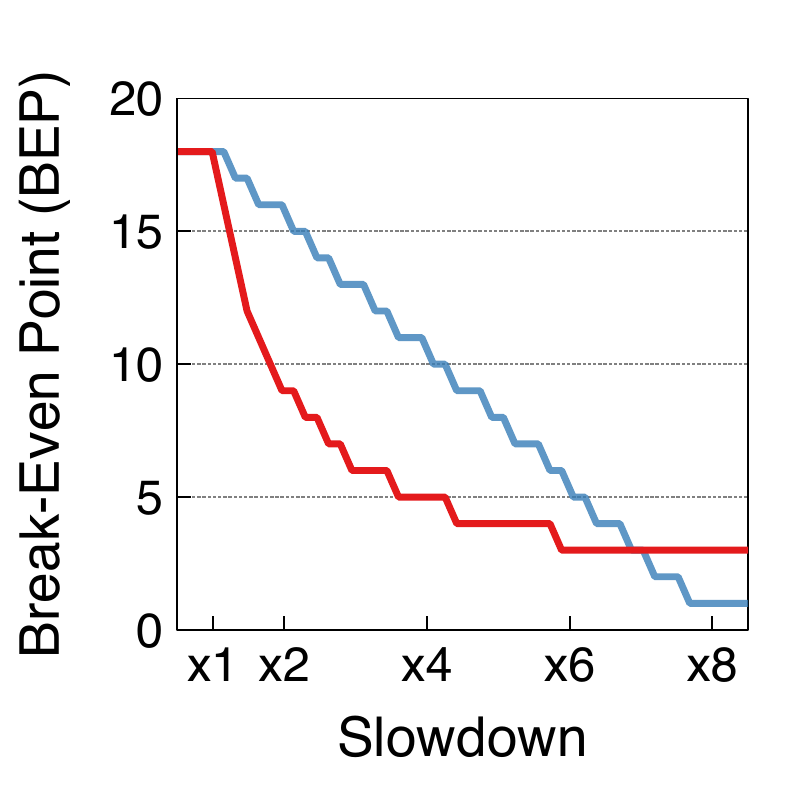} &
       \includegraphics[width=0.23\linewidth]{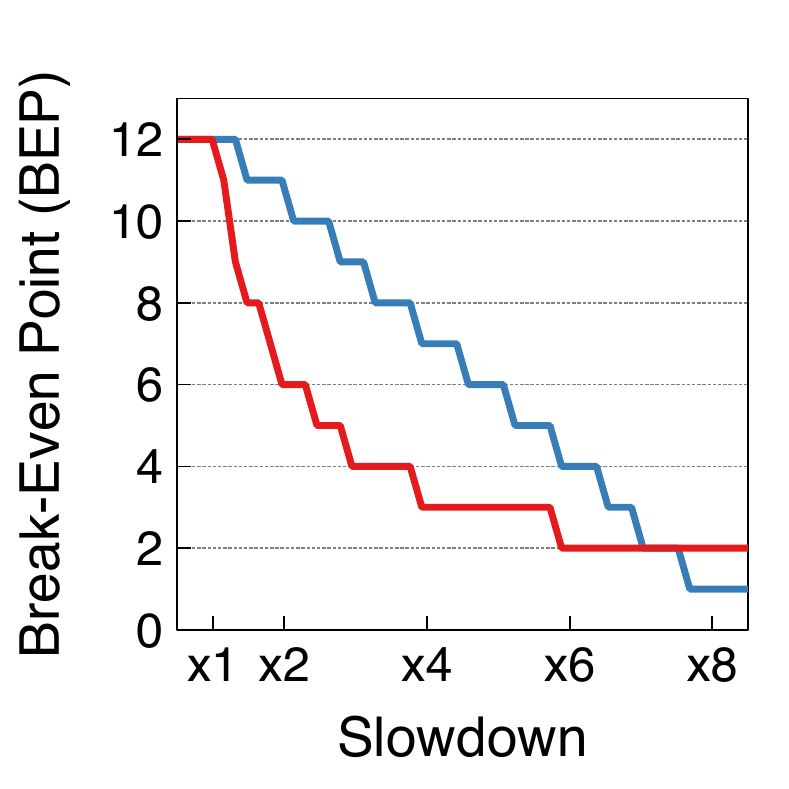} &
       \includegraphics[width=0.23\linewidth]{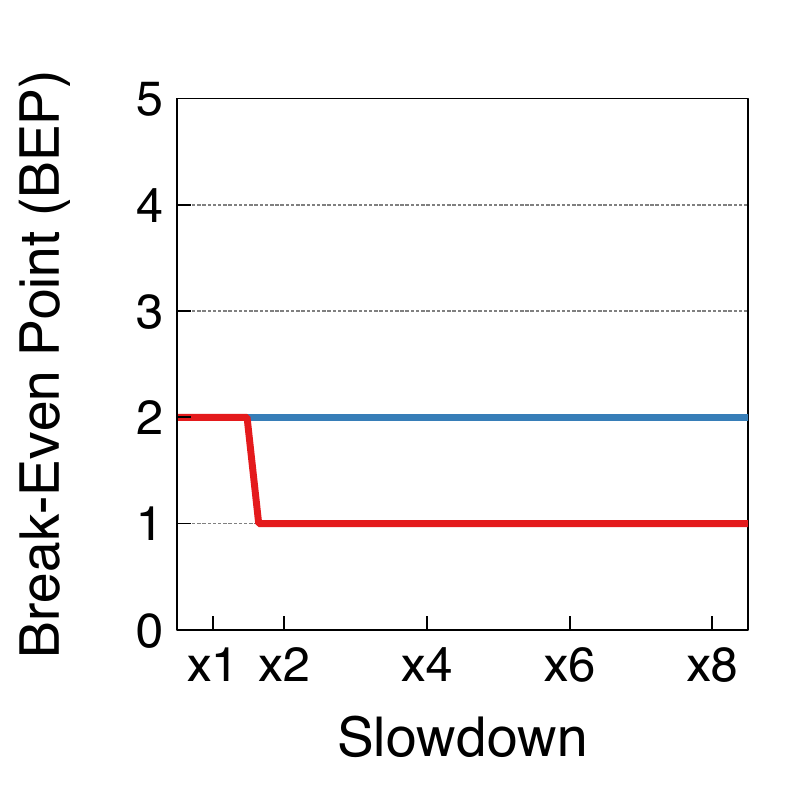}\\
       \vspace{-5pt}
       (e) Count & 
       (f) Vector Addition &
       (g) Array Merge &
       (h) Page Rank\\
       \multicolumn{4}{c}{Newport CSD-array system}\\
    \end{tabular}
	\caption{
	{
    	Analysis of reduction in the number of devices (BEP) by varying host's slow-down factors.
    	} 
    }
    \vspace{-10pt}
	\label{plot:model_proj_overload}
\end{figure*}

\begin{figure}[!t]
	\centering
    \includegraphics[width=0.5\linewidth]{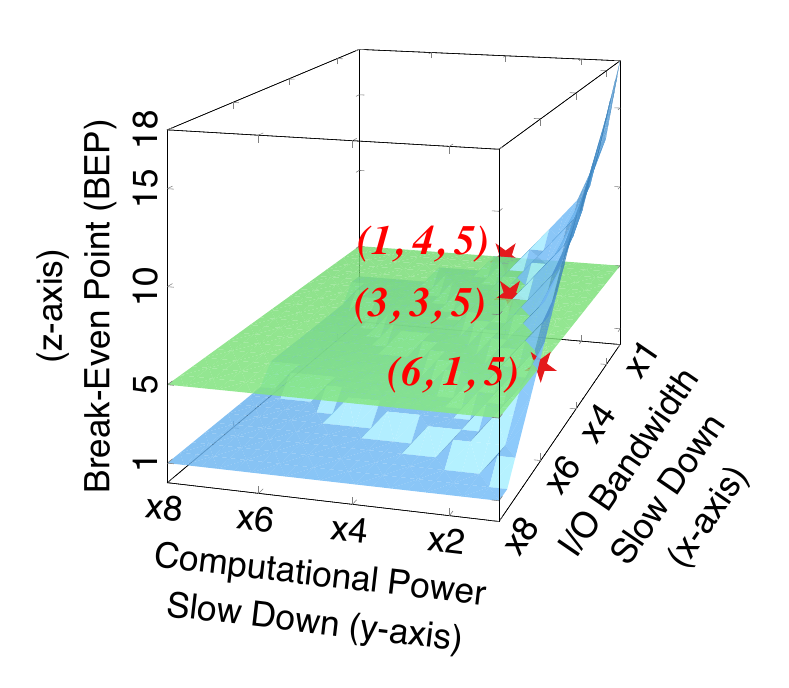}\\
	\caption{
	{Finding the values of BEP according to the change of host's slow-down factors for Vector Addition with the Newport CSD-array.
	}
	}
    \vspace{-5pt}
	\label{plot:model_overload_change_3d}
\end{figure}

Figure~\ref{plot:model_proj_overload}(a)-(d) shows the results for SmartSSD. The result is similar to Figure~\ref{plot:model_proj_normal}(a)-(d). In this evaluation, the decrease in the computational power of the host almost coincides with the increase in the computational power of the SmartSSD in Figure~\ref{plot:model_proj_normal} (red line).
On the other hand, for Vector Addition and Array Merge in Figure~\ref{plot:model_proj_normal}, the increase in SmartSSD's internal I/O bandwidth has a limit to the BEP reduction, but the decrease in the host's I/O bandwidth has an almost linear effect on the BEP.
In Figure~\ref{plot:model_proj_normal}, since BEP reduction is evaluated when the total execution time of the SmartSSD is reduced, the total execution time of the workload is bound to the computation time.
However, in this evaluation, since BEP reduction was evaluated when the total execution time of the host's workload increases, the total execution time of the workload continues to increase according to the decrease in the I/O bandwidth of the host.
Page Rank is completely compute-intensive, so BEP is not affected by the reduced host's I/O bandwidth.

Figure~\ref{plot:model_proj_overload}(e)-(h) shows the results for Newport CSD. Unlike SmartSSD, this result is not similar to Figure~\ref{plot:model_proj_normal}(e)-(h). This is because Newport CSD has lower I/O bandwidth and computing power compared to the host. Note that the host system uses SmartSSD as a block device, so the I/O bandwidth is similar to SmartSSD.
Overall, BEP is exponentially affected by the reduction in host computational power, while it is linearly affected by the reduction in I/O bandwidth. The reason is that, in Equation~(\ref{eq:model_func_overloaded}), the value of the $S_\mathrm{{overload}}$ function is inversely proportional to $sd_{\mathrm{comp.}}$, whereas it is in direct proportional relationship to $sd_{\mathrm{tx}}$.
In Count, Vector Addition, and Array Merge, BEP is more affected by the reduction in computational power than the reduction in I/O bandwidth of the host and reverses after $\times$7 in Figure~\ref{plot:model_proj_overload} 
The inversion value is dependent on the performance difference between CSD, host, and workload characteristics.
Page Rank is completely compute-intensive, so BEP is not affected by the reduced host's I/O bandwidth.

In addition, \cplan{} finds BEP for the combination of two factors ($sd_\mathrm{tx}$, $sd_\mathrm{comp.}$), as shown in Figure~\ref{plot:model_normal_change_3d}.
Figure~\ref{plot:model_overload_change_3d} visually shows the BEP values that \cplan{} finds. The $x$-axis means I/O bandwidth ($sd_\mathrm{tx}$), and the $y$-axis means computational power ($sd_\mathrm{comp.}$), and the $z$-axis means BEP (the number of devices) that \cplan{} finds.
In the figure, the explanation of BEP is the same as that of Figure~\ref{plot:model_normal_change_3d}.
For example, sketch red stars.

\begin{figure}[!t]
	\centering
    \includegraphics[width=0.5\linewidth]{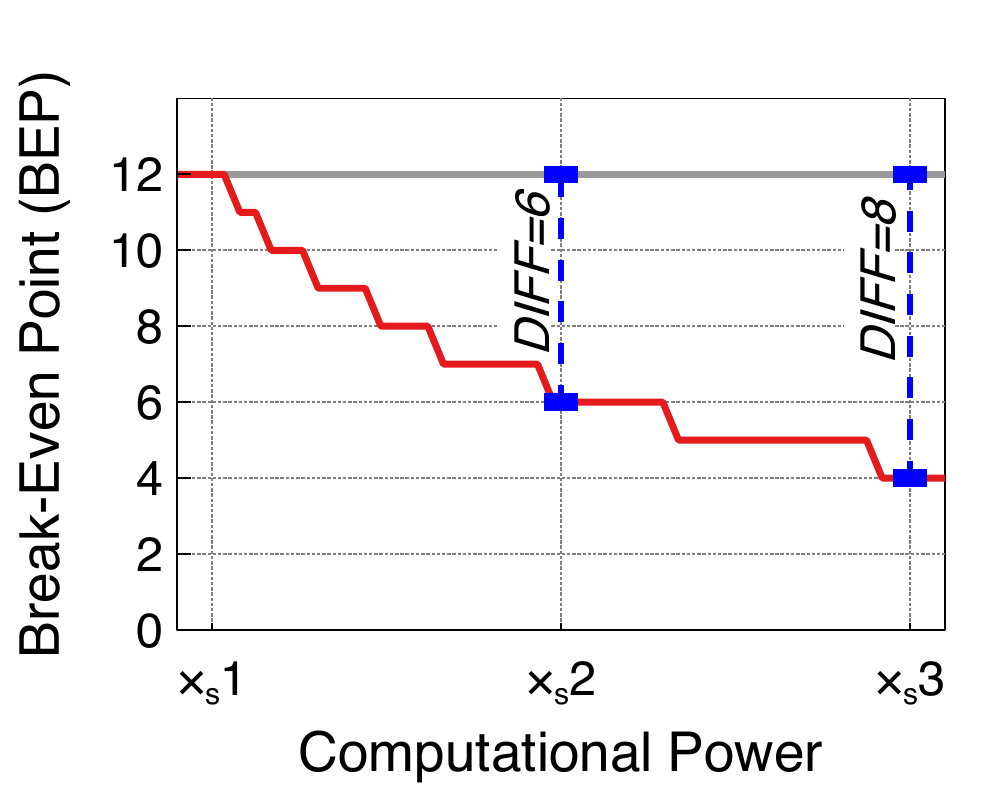}\\
	\caption{
	{{In Array Merge of Newport CSD, BEP reduction is according to the computational power of the host CPU. On the $x$-axis, '$n$' in $\times_{s}n$ indicates $n$ times slower.}
	}
	}
    \vspace{-5pt}
	\label{plot:change_bep_device}
\end{figure}

\subsection{{Analysis of Total Cost of Ownership}}

{
In this subsection, we will discuss the total cost of ownership for building a compute node using \cplan{}.
The cost of building a server consists of installation costs and recurring costs. We compare the cost of building a compute node using SSDs and CSDs where we only consider the cost of host CPU and storage among the installation costs and ignore the recurring cost. Also, we consider a system with the same throughput for processing workloads for both systems. CSD-system can improve throughput through internal resources, so using a low computational power CPU for the host would reduce the cost.
}

{
Figure~\ref{plot:change_bep_device}, in Array Merge of Newport CSD, shows the difference (\textit{DIFF}) between the BEP and the initial BEP value (12) according to the decrease in computational power of the host CPU.
The $x$-axis represents the slowdown factor of computational power, and the \textit{DIFF} represents the difference in the number of CSDs required to meet the performance requirement of the SSD system for the CSD system. For instance, the CSD-system with $2\times$ slower host CPU only requires 6 Newport CSDs to meet the performance requirements, while CSD-system with $3\times$ slower host CPU only requires 8 Newport CSDs. 
}

\renewcommand{\arraystretch}{1.3}
\begin{table}[t]
\centering
    \caption{{The Total cost of ownership when building a compute node with baseline SSDs and CSD. }
    }
	\vspace{-10pt}
    \resizebox{0.93\textwidth}{!}{
\begin{tabular}{|c||cccc||cc||c|c|c|}
\hline
\multirow{2}{*}{System} & \multicolumn{4}{c||}{\textbf{CPU-Part}}                                                                  & \multicolumn{2}{c||}{\textbf{Storage-Part}} & \multirow{2}{*}{\textbf{\begin{tabular}[c]{@{}c@{}}Total\\ Cost\end{tabular}}} & \multirow{2}{*}{{\textbf{\begin{tabular}[c]{@{}c@{}}Mark / \\ Dollar\end{tabular}}}} & \multirow{2}{*}{\textbf{\begin{tabular}[c]{@{}c@{}}Saved\\ Cost\end{tabular}}} \\ \cline{2-7}
                        & \multicolumn{1}{c|}{Name}          & \multicolumn{1}{c|}{Mark}  & \multicolumn{1}{c|}{Slow Down} & Cost & \multicolumn{1}{c|}{SSD/CSD}     & Cost    &                                                                                &                                                                                  &                                                                                \\ \hline\hline
\textbf{SSD-system}     & \multicolumn{1}{c|}{AMD EPYC 7351} & \multicolumn{1}{c|}{39999} & \multicolumn{1}{c|}{-}         & \$1828 & \multicolumn{1}{c|}{12$\times$SSD}      & \$1200    & \$3028                                                                           & {13.2}                                                                            &  -                                                                              \\ \hline
\textbf{CSD-system (A)}  & \multicolumn{1}{c|}{AMD EPYC 7452} & \multicolumn{1}{c|}{20471} & \multicolumn{1}{c|}{Appx. $\times$2}    & \$1399 & \multicolumn{1}{c|}{6$\times$CSD}       & \$660     & \$2059                                                                           & {19.4}                                                                            & 32\%                                                                             \\ \hline
\textbf{CSD-system (B)}  & \multicolumn{1}{c|}{AMD EPYC 7251} & \multicolumn{1}{c|}{14935} & \multicolumn{1}{c|}{Appx. $\times$3}    & \$485  & \multicolumn{1}{c|}{8$\times$CSD}       & \$880     & \$1365                                                                           & {29.3}                                                                            & 35\%                                                                             \\ \hline
\end{tabular}
}
\label{tbl:cost_save} 
\end{table}

{Table~\ref{tbl:cost_save} shows an example of a cost comparison between the SSD-system and CSD-system (A and B) {related to Figure~\ref{plot:change_bep_device}}.  
The first part of Table~\ref{tbl:cost_save} represents the adopted CPUs in all three systems. We considered a system with AMD EPYC 7351 and 12 SSDs as the baseline compute node. For CSD-systems, we considered the low-power CPUs from the same lineup, AMD EPYC 7452 (approximately $2\times$ slower) and AMD EPYC 7251 (approximately $3\times$ slower), as CSDs have considerable computation resources~\cite{cpumark}. The second part of Table~\ref{tbl:cost_save} shows the storage devices employed in all three systems. The SSD-system is equipped with 12 SSD while CSD-system (A) and CSD-system (B) have a varying number of CSDs based on the BEP values from \cplan{}. 
}
{
The commercially available CSDs are relatively expensive compared to SSDs. Thus, we assume that the prices of CSDs would become affordable once adopted by the system architect actively. For this comparison, we consider the following cost model for SSD and CSD: 
(i) The price of the SSD is \$100, and (ii) The cost of processor-equipped SoC for CSD is 10\% higher, so the price of CSD is \$110. 
Comparing the total cost according to our model, CSD-system (A) and CSD-system (B) can reduce costs by 32\% and 55\% compared to SSD-system, respectively.
In other words, system architects can use \cplan{} to reduce server building costs by utilizing CSD according to the usage environment of the compute node.
\section{Conclusion and future work}
\label{sec:conc}

HPC facilities have started looking at the potential of adopting storage devices within the simulation nodes which provides an opportunity for adopting in-storage processing solutions within simulation nodes to perform data analysis tasks. With the advent of CSDs, there are opportunities for building CSD-array-based computing nodes called \cstore{}. With CSDs, data analytic tasks are offloaded to the device where data resides and, reducing the cost of data movement optimizing the performance, energy utilization, and total cost of ownership. However, adopting CSDs naively does not benefit due to the distinct hardware and performance characteristics of commercially available CSDs. Therefore, in this work, we formulated and implemented a storage capacity planner, called \cplan{}, that takes into account the hardware and performance characteristics of CSDs, host systems, and workloads to provision \cstore{} in a cost-effective manner. \cplan{} finds the optimal number of CSDs (BEP) to outperform a traditional compute node with block-based SSDs. We demonstrated the efficacy of our proposed \cplan{} through two commercial CSDs -- SmartSSD and Newport CSD -- and showed how \cplan{} effectively finds optimal BEP. Our proposed solution also tracks changes in BEP according to the change in hardware parameters of host and CSD systems (i.e., computational power and I/O bandwidth). 

{The simulation node can adopt a CSD and SSD combination system (CSD-SSD system). In this case, \cplan{}'s capacity planner alone is not sufficient to find the optimal BEP according to the workload. In the CSD-SSD system, the degree of performance improvement due to parallel processing varies according to the number of SSDs and CSDs. In addition, the size of the workload executed by the CSD-SSD system can be dynamically changed depending on the situation, and several different workloads can be executed simultaneously. Therefore, in this case, sophisticated workload analysis is required considering the number of SSDs as well as the performance characteristics of CSDs. We will expand \cplan{} as future work to explore technologies that allow CSD-SSD systems to have optimal performance in dynamic workloads.}

\bibliographystyle{ACM-Reference-Format}
\bibliography{main}

\end{document}